# Polariton-Based Room Temperature Quantum Phototransistors

Jhuma Dutta[1], Pooja Bhatt[1], Kuljeet Kaur[1], Daniel E. Gómez[2]* and Jino George[1]*

*[1]Indian Institute of Science Education and Research (IISER) Mohali, Punjab-140306, India.*
*[2]School of Science, RMIT University, Melbourne, VIC, 3000, Australia.*

Corresponding authors: jgeorge@iisermohali.ac.in; daniel.gomez@rmit.edu.au

**Abstract: Strong light-matter coupling is a quantum process in which light and matter are coupled together, generating hybridized states. This is similar to the notion of molecular hybridization, but one of the components is light. Here, we utilized the idea and prepared quantum phototransistors using donor-acceptor combinations that can transfer energy via Rabi oscillations. As a prototype experiment, we used a cyanine J-aggregate (TDBC; donor) and MoS2 monolayer (acceptor) in a field effect transistor cavity and studied the photoresponsivity. The energy migrates through the newly formed polaritonic ladder, and the relative efficiency of the device is nearly seven-fold at the ON resonance. Further, the photon mixing fraction is calculated for each independent device and correlated with energy transfer efficiency. In the strongly coupled system, newly formed polaritonic states reshuffle the probability function. A theoretical model based on the time dependent Schrödinger equation is also used to interpret the results. Here, the entangled light-matter states act as a strong channel for funnelling the energy to the MoS2 monolayer, thereby boosting its ability to show the highest photoresponsivity at ON-resonance. These experimental findings and the proposed model suggest novel applications of strong light-matter coupling in quantum materials.**

## Introduction:

Recent developments in the miniaturization of devices are achieved by introducing photonic components. The virtual memory of the photonic states is utilized to prepare fast-responding quantum devices.[1] Further, photonic and excitonic states can be mixed through strong light-matter interaction, generating quasi-Bosonic particles known as polaritons.[2] Here, the transfer of energy/charge is much more efficient as polaritonic states have a small effective mass, still inheriting material behavior. Such a system shows interesting phenomena such as Bose-Einstein condensation[3], polariton lasing [4] etc. Cavity polaritons have emerged as a novel foundation for optical devices, showcasing numerous compelling concepts for all-optical integrated logical circuits[5]. Recent experimental achievements include the demonstration of polariton spin switches[6], polariton transistors[7], and polariton interferometers[8]. The same idea can be translated into molecular systems by mixing their transition dipole moment with the electric dipole of the confined light.

Transfer of energy and/or charge are some of the most fundamental processes in nature, and understanding and manipulating these processes is intriguing.[9] For example, plasmon-coupled energy transfer is efficient in hybrid nanoparticle systems.[10] There is experimental evidence for long-range dipole-dipole interactions facilitated by surface lattice resonances, which help delocalize excited-state energy over long distances.[11] Another way to increase the longitudinal radiationless transfer is through the modification of the photonic density of states by introducing a mirror in front of an emitter.[12] This can further affect the energy transfer (ET) efficiency between a D-A pair.[13] Notably, most of these studies occur in the so-called weak coupling regime. This situation changes when the energy exchange rate between light and matter is faster, forming entangled light-matter states. In 2015, Ebbesen's group demonstrated that material properties of organic semiconductors, such as conductivity and photoconductivity, can be enhanced by injecting charges into the polaritonic states.[14,15]

Further, a few theoretical investigations suggest that the phenomenon of polariton transport can substantially enhance the conductivity of organic materials.[16,17] Strong light-matter interaction has emerged as a tool for manipulating material properties in a somewhat unforeseen manner.[18] For instance, the new excited states offer access to new ultrafast energy relaxation pathways [19,20] and exhibit modified work functions [21] and chemical reactivity[22], characteristics that usually arise from the hybrid light-matter states. Moreover, the emergence of polaritonic states can lead to unusual modifications of energy transfer processes.[23,24] In this work, we emphasize that strong coupling significantly changes the ET process through the formation of polaritonic states that help us to build a quantum optoelectronic device at room temperature.

In polariton-mediated ET, probability function is affected due to the formation of new states. Very interestingly, polaritonic states act as an efficient medium for energy migration.[25,26] Long-range ET between spatially separated entangled D-A systems shows that strong coupling surpasses the limitations of the Förster distance, and it is now defined by the mode volume of the polaritonic states.[27-29] This is particularly achieved by the domination of transverse long-range components.[30] Motivated by these experiments, a cavity quantum electrodynamics concept has been developed for polariton-assisted ET in strongly coupled systems that can study the energy transfer beyond conventional Förster distance.[9,30] For example, ballistic exciton transport is reported for a strongly coupled system that can allow energy to funnel through the polaritonic states much more efficiently.[31,32] More information on polariton-mediated energy transfer can be obtained from the suggested review article.[23] A similar idea is applied in quantum cascade detectors by preparing light-matter coupled states in the infrared regime.[33]

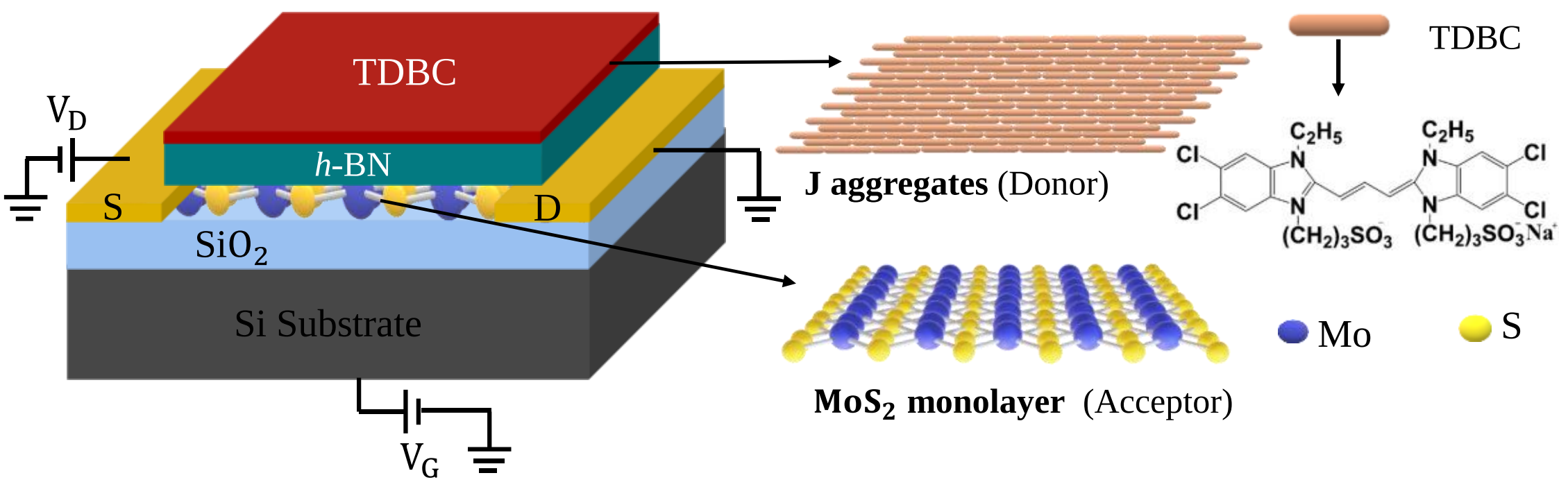

***Figure 1.*** Schematic representation of a D-A phototransistor: The device (left panel) composed of TDBC J-aggregate (donor), and a $MoS_2$ monolayer (acceptor); (right panel) the molecular structure of the organic dye TDBC and a head-to-tail orientation of the J-aggregate, and (right-bottom) a three-dimensional representation of the structure of $MoS_2$ monolayer.

There are many studies on the energy/electron transport in strongly coupled systems.[34] However, these studies are limited to spectroscopic demonstration of the idea. This work proposes a new quantum device based on polaritonic states, in which the energy transport phenomenon is studied through photoconductivity measurements using an organic-inorganic hybrid interface. The combination of organic and inorganic semiconductors to create a hybrid heterojunction presents a distinctive opportunity to leverage the advantages of both materials. For instance, TDBC J-aggregates are well-known for their strong light absorption capabilities [35] but have poor electron mobility, whereas $MoS_2$ exhibits a relatively low absorption coefficient, yet excels in charge transport characteristics.[36,37] Here, we used a well-studied D-A pair for cavity experiments.[38]

Quasi-particles like polaritons behave collectively and interact with each other through photonic degrees of freedom; they play a crucial role in the developing the field of polaritronics.[39] In this work, we employ photoconductivity measurements in a metal-oxide-semiconductor field-effect transistor (MOSFET) configuration to investigate the ET from TDBC (donor) to an atomically-thin $MoS_2$ monolayer (acceptor). We demonstrate ET between donor-acceptor layers separated by as much as 1500 Å, mediated through polaritonic states in a Fabry-Pérot (FP) cavity. Evidence of ET is obtained by mapping the photoresponsivity of the $MoS_2$ monolayer in the presence of the donor through cavity detuning experiments. A model based on a time-dependent Schrödinger equation with a Hamiltonian that accounts for the interaction among three subsystems: a cavity mode, donor and acceptor show good agreement with experiments, suggesting ET occurs through Rabi oscillations. Further, the energy transfer probability function can be tailored to achieve a desirable quantum phototransistor without modifying the active layer.

## Results and Discussion

We compare the performance of conventional and polaritonic state-based phototransistors. We chose a strong absorber (J-aggregate dye) which acts like an antenna and transfers its energy to an atomically thin $MoS_2$ monolayer. ET experiments were conducted on two configurations: (1) at D-A separation distances where ET occurs *via* short-range dipole-dipole interactions (i.e. D-A distances of about less than 100 Å) and (2) radiative polariton-mediated long-range ET in a cavity system, by taking advantage of the extended excited-state delocalization of the D-A pair under strong coupling. Further, we create a hybrid phototransistor utilizing an organic-inorganic heterojunction, which serves as a platform for investigating the ET process across the interface. A three-terminal gated configuration was created to analyze the electrical characteristics of the hybrid system, as depicted schematically on the left side of **Figure 1**. The phototransistor comprises a heterojunction having a layer-by-layer (LBL) assembly of TDBC J-aggregates and $MoS_2$ monolayer separated by a hexagonal boron nitride (*h*-BN) dielectric spacer. The chemical structure of the organic dye TDBC and the $MoS_2$ monolayer are on the right panel of **Figure 1**. J-aggregate formation in the organic dye arises from the dipole alignment (head-to-tail) of TDBC molecules.[40] TDBC J-aggregate thin films are deposited on PDMS substrate using the LBL assembly technique as explained in the Methods section (SI).[40] The LBL approach ensures precise control over the film thickness at the nanometer scale and has high absorption cross-sections with low optical scattering (homogeneous). We have deposited six layers of TDBC corresponding to an approximate film thickness of 10 nm.

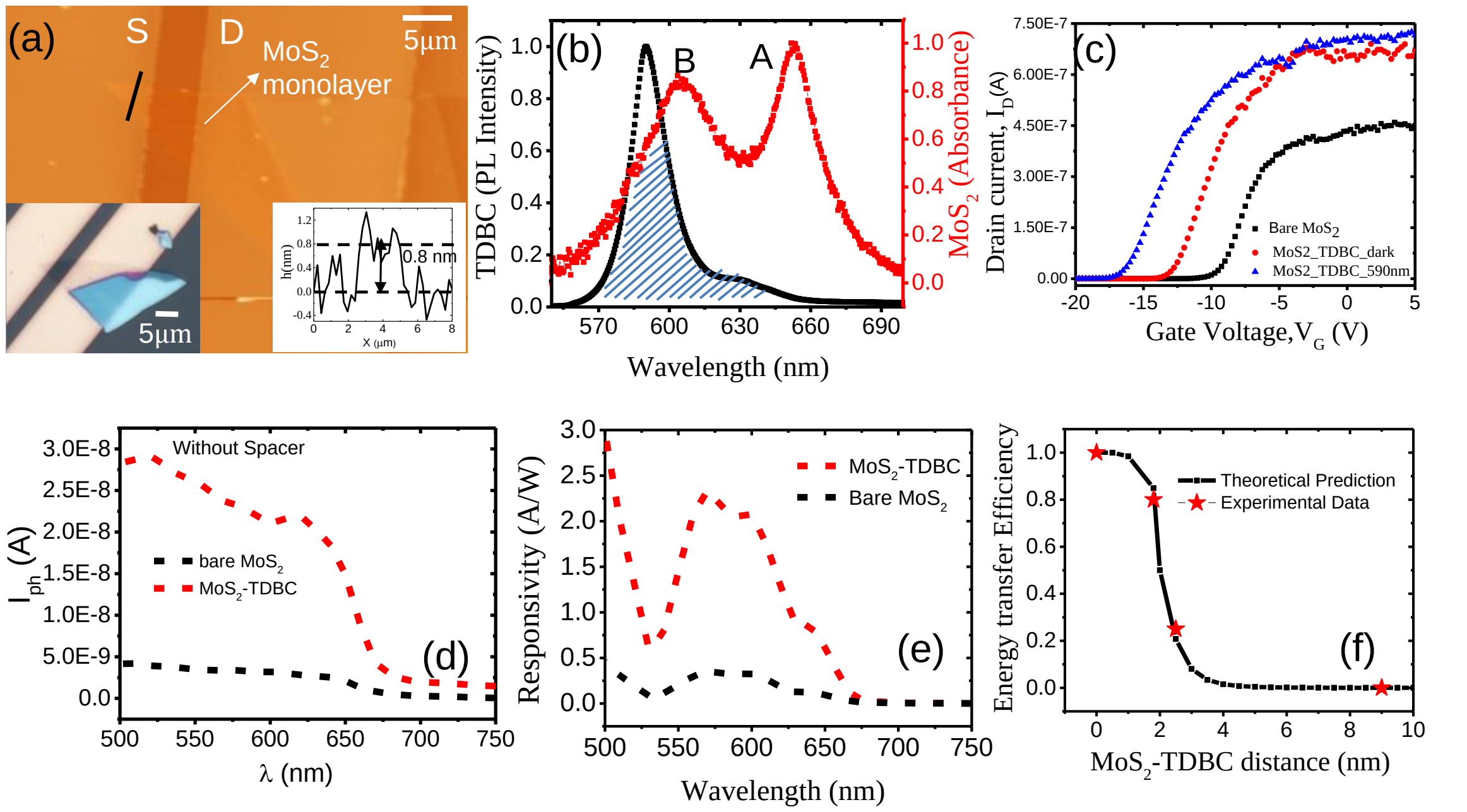

***Figure 2.*** Non-cavity experiments: (a) AFM image of $MoS_2$ monolayer on the MOSFET device having a thickness of ~0.8 nm (Inset: AFM height analysis). Inset on the left: optical image of the $MoS_2$ monolayer (S-source; D-drain; Ti/Au electrode with L=5 µm and W= 5.5 µm). (b) Photoluminescence emission spectra arising from the TDBC J-aggregate, along with the absorption spectra of a monolayer of $MoS_2$. The region of spectral overlap is depicted through shading. (c) $I_D$-$V_G$ characteristics of bare $MoS_2$, $MoS_2$/TDBC at dark and at an excitation wavelength of 590 nm, where the source - drain bias voltage is at 0.5 V. (d) Photocurrent spectra of $MoS_2$ monolayer, before (black) and after (red) TDBC deposition without any spacer layer. (e) Photoresponsivity as a function of laser excitation wavelength before (black) and after (red) TDBC deposition ($V_D$=0.5V, $V_G$=-9.5V). (f) ET efficiency as a function of separation distance (in nm) between $MoS_2$ monolayer and TDBC thin film. The theoretical expectation, following resonance ET theory, is represented by the continuous line.

A monolayer of $MoS_2$ is mechanically exfoliated using the scotch tape method on thin PDMS stamps, and later transferred to a gold electrode as shown in the inset of **Figure 2a**. The presence of a monolayer of $MoS_2$ was confirmed by both AFM and photoluminescence measurements.[41] The thickness of the multilayer *h*-BN was measured by AFM height analysis as shown in (Figure S4). The *h*-BN layer is a two-dimensional insulator with a single-layer thickness of ~ 0.4 nm and a bandgap of ~ 6 eV.[42] On the other hand, the $MoS_2$ monolayer has an approximate thickness of 0.8 nm with a direct bandgap of 1.9 eV.[41] The size of the monolayer flakes is roughly 5-6 μm. **Figure 2b** displays the PL spectrum of TDBC J-aggregates (black solid lines) and absorption spectrum of $MoS_2$ monolayer (red solid lines) where the shaded region highlights their spectral overlap. PL emission of TDBC occurs at 590 nm and the absorption spectrum of the $MoS_2$ monolayer exhibits two distinct Wannier-Mott exciton resonances: an A-exciton at 655 nm, and B-exciton at 606 nm. The near-perfect spectral overlap between the TDBC PL emission and the $MoS_2$ B-exciton absorption establishes an optimal condition for ET to occur. Selective excitation was done at 532 nm using a diode laser and monitoring the PL spectra of TDBC (emission maximum at 590 nm) quantifies ET in the D-A system (Figure S1).

We used photoresponsivity spectra (PS) for mapping the ET process to $MoS_2$ by measuring the photoconductivity as a function of wavelength. $MoS_2$ demonstrates excellent photoconductivity and exhibits a clear ON-OFF ratio within a three-terminal configuration.[43] Electrical characteristics ($I_D$-$V_G$ curve; $I_D$: drain current and $V_G$: gate voltage) of bare $MoS_2$, $MoS_2$-TDBC at dark and at an excitation wavelength of 590 nm are shown in **Figure2c**, with a source-drain bias voltage of 0.5 V. A change in the threshold voltage ($\Delta V_T$) is also observed across our experimental measurements. There are two possible ways to affect the threshold voltage; (i) by introducing a dielectric layer on top of the active layer, and/or (ii) by photo-illumination.[44] These details will be elaborated while discussing the cavity data. Subsequently, the photocurrent ($I_{ph}=I_{illumination}-I_{dark}$) spectra of $MoS_2$ monolayer, before (black) and after (red) TDBC deposition was acquired and it is shown in **Figure 2d.** These spectra were obtained by illuminating the active region of the device using a supercontinuum laser possessing a variable filter with a 10 nm bandwidth.[45] Please note that the photoresponsivity ($I_{ph}$ /P; P is the illumination power) gives a normalized response of the photodetector.

The photoresponse of the intrinsic $MoS_2$ devices was measured across the visible range (500-750 nm) at room temperature. The photocurrent of bare $MoS_2$ starts increasing at its band edge (~ 670 nm) and then continuously increases with decreasing excitation wavelength, and saturates at higher energies. Similar measurements were performed in the D-A system without a spacer layer. The photoresponsivity spectra (PS) are compared for bare $MoS_2$ and TDBC-$MoS_2$ in **Figure 2e,** where $V_G$ = -9.5 V. In the D-A system, the enhancement in responsivity was noted to follow the donor-acceptor spectral overlap function. When the *h*-BN spacer is inserted between TDBC and $MoS_2$, the energy transfer channel is still accessible (Figure S3), and the measured ET efficiency beautifully follows the predictions of standard resonance ET theory. **Figure 2f** shows the variation of the ET efficiency as a function of the separation between TDBC and $MoS_2$; it abruptly decreases for distances greater than 100 Å. The detailed calculation of energy transfer efficiency from photoresponsivity data is given in section 4 and the estimated Förster radius was found to be 20 Å for the D-A pair (**Figure 2f**). When the spacer layer increases beyond 100 Å, ET ceases completely for the D-A pair in this configuration. This suggest that pure short-range dipole-dipole interaction is taking place in the non-cavity experiments.

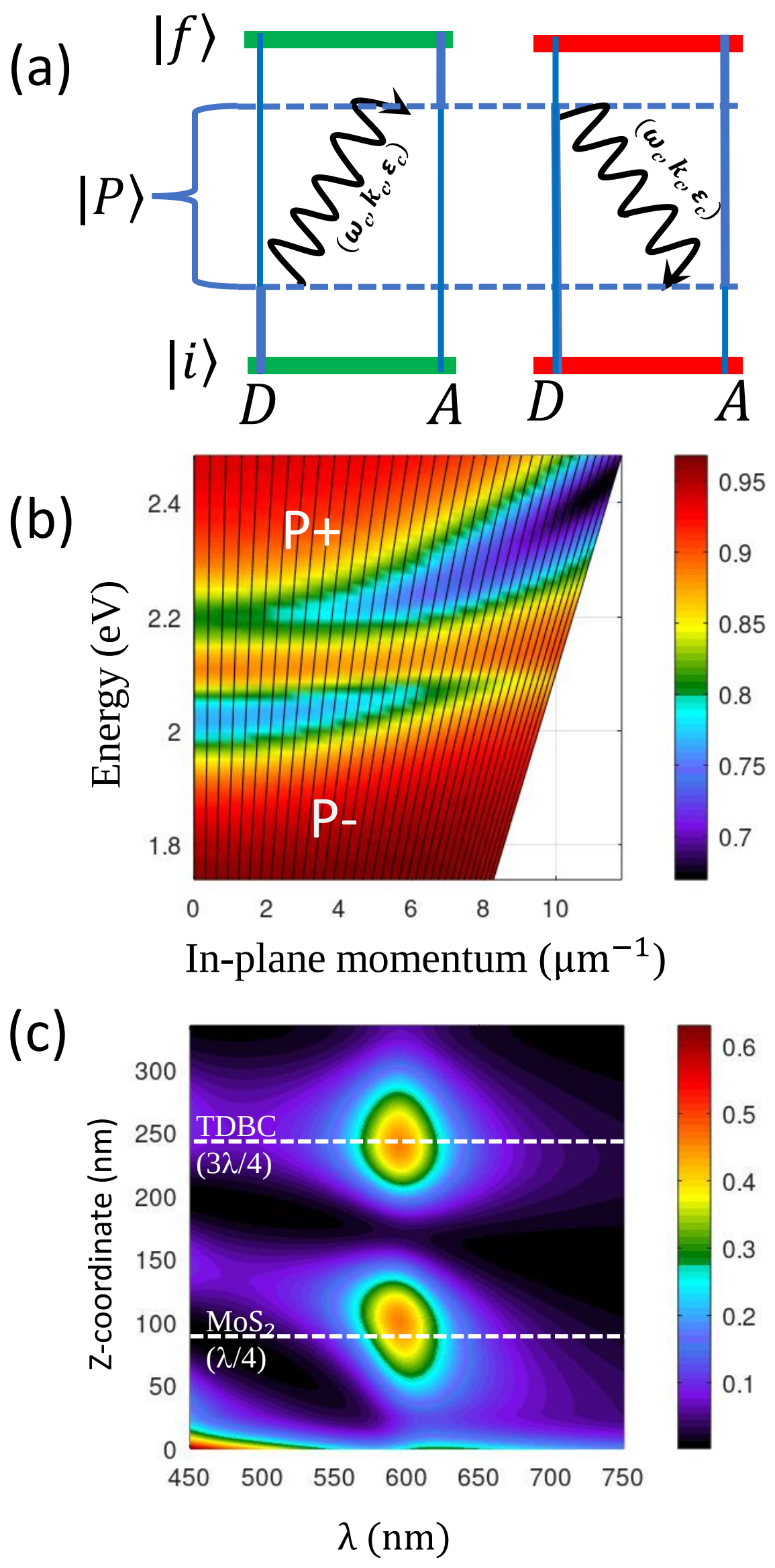

***Figure 3.*** *Cavity - optical output of a quantum phototransistor*: (a) Schematic representation showing spatially separated, radiative polariton-mediated energy transfer processes in strongly coupled D-A system. (b) TMM calculated energy/in-plane momentum dispersion of polaritonic states. (c) Electromagnetic field distribution, computed using TMM shows the formation of λ mode; both donor (TDBC) and acceptor ($MoS_2$) are placed at the antinode positions, mentioned as white dotted lines.

Strong light-matter coupling presents an alternative approach for energy transfer that extends beyond Förster distance. This approach offers an additional benefit of customizing the spectral overlap function between the donor and the acceptor components. Our earlier attempts on long-range energy transfer by switching the $MoS_2$ to the donor position clearly show that the energy can be transferred to an acceptor molecule under cooperative coupling conditions. Cooperative coupling of spatially separated excitons can efficiently transfer energy as long as the cavity is coupled to the D-A system.[43] In this work, the donor TDBC and the acceptor $MoS_2$ are spatially separated from each other. Since TDBC has a high oscillator strength, only the donor is strongly coupled to the cavity, forming polaritonic states. On the other hand, the acceptor (monolayer $MoS_2$) exhibits weak coupling due to its low oscillator strength. **Figure 3a** represents dipole-allowed, radiative polariton-mediated energy transfer process between an initial state $|i\rangle$ and a final state $|f\rangle$. These processes involve polaritonic states $|P\rangle$, in which a cavity photon with frequency $\omega_c$, wavevector $k_c$, and polarization $\varepsilon_c$ is exchanged. The energy/in-plane momentum dispersion for the FP cavity simulated by transfer matrix method (TMM) calculations confirm the formation of two polariton branches (**Figure 3b**). The formation of polaritonic states opens new radiative channels in the hybrid system that facilitates an efficient ET process.

MOSFET-Fabry-Pérot (MOSFET-FP) cavity devices were prepared with a fixed *h*-BN spacer layer thickness of 1500 Å as shown in **Figure 4a**. Cavity structures were fabricated with a silicon substrate on one side, leveraging its high dielectric constant, while the opposite side featured a Ag mirror (~30 nm thick) serving as a reflective surface. To complete the cavity configuration, thin films of TDBC J-aggregates were deposited on a silver-coated PDMS substrate by LBL assembly technique and then placed on top of the silicon/$MoS_2$/ *h*-BN substrate. In order to maximize exciton-photon coupling, the thickness of the PMMA layer was adjusted such that two active layers (separated by *h*-BN spacer) were strategically positioned at the points of maximum standing wave intensity within the full wavelength (λ) cavity. An electric field intensity map was calculated via a TMM simulation assuming a cavity length of 300 nm, with TDBC and $MoS_2$ layers located at the antinode positions (λ/4 for $MoS_2$ and 3λ/4 for TDBC; **Figure 3c**).

This structural configuration proves advantageous for investigating the field effect properties of semiconductors under strong light-matter coupling. The polaritonic states are formed from a single mirror cavity configuration, which results in a low-quality factor of 5-6, evidenced by a broad spectral linewidth (**Figure 4b**). In the presence of the *h*-BN spacer, the photocurrent of the $MoS_2$ monolayer fluctuates, and hence the “non-cavity” reference has been taken as $MoS_2$/*h*-BN. This approach ensures that any alterations in current readings attributed to *h*-BN are accounted for in advance. Figure S7 depicts the photoresponsivity as a function of the excitation wavelength for the cavity system. A comparison to the non-cavity photoresponsivity data indicates that there is no ET if the *h*-BN spacer layer is 1500 Å. However, upon introducing the $MoS_2$/ *h*-BN/TDBC composite into the cavity, a large enhancement is evident, as illustrated by the red dotted lines in Figure S7. For the control experiment, the $MoS_2$/ *h*-BN alone was placed inside the cavity, but no change in PS was observed (Figure S6). This observation emphasizes the significance of strong coupling to control the ET process. To confirm that the enhancement in photoresponsivity is a consequence of polariton-mediated effects, we conducted photocurrent measurements in several devices with differing PMMA thickness, therefore allowing us monitor both optical and electrical outputs by cavity tuning experiments.

**Figure 4c** illustrates the ratio between the photoresponsivity of the cavity and non-cavity ($MoS_2$/*h*-BN) at 590 nm excitation wavelength, plotted against the estimated empty cavity mode energy (e.g. wavelength). Both PS and the optical reflection spectra corresponding to various cavity energies are given in the SI (Section 9). **Figure 4c** reveals that the most significant enhancement in PS occurs when the cavity mode energy is ON- resonance with the excitons in the J-aggregates. The calculated dispersion spectra as a function of resonance cavity mode position are shown in **Figure 4d**, where the pink and green solid lines represent the energy of P+ and P- states, respectively. These polariton branches were simulated and analysed using TMM calculations. To establish a connection between the optical and electrical outcomes of the devices, we plotted the experimental dispersions of both the P+ and P- states against the position of the empty cavity mode. The experimental P+ and P- positions are denoted by blue and red solid spheres, respectively, and exhibit a Rabi splitting ($\hbar\Omega_R$) energy of 179 meV. The observed $\hbar\Omega_R$ follows strong coupling criteria, where $\Omega_R > \Gamma_c, \Gamma_m$, where $\Gamma_c$ (150 meV) and $\Gamma_m$ (35 meV) are the decay constants of the cavity and donor molecules, respectively (Section S13). The photoresponsivity enhancement due to ET at the ON-resonance position is now correlated with the spectral measurements. Even though, the cavity has a huge dissipation loss (due to its low-quality factor), a strong coupling condition can be achieved with a one-mirror configuration. Now, the newly formed polaritonic branches allow energy to spatially delocalize through the photonic cavity mode and reach the acceptor molecule kept at a distance within the cavity. The question that arises is: what are the deciding factors controlling the energy flow between the entangled D-A pair?

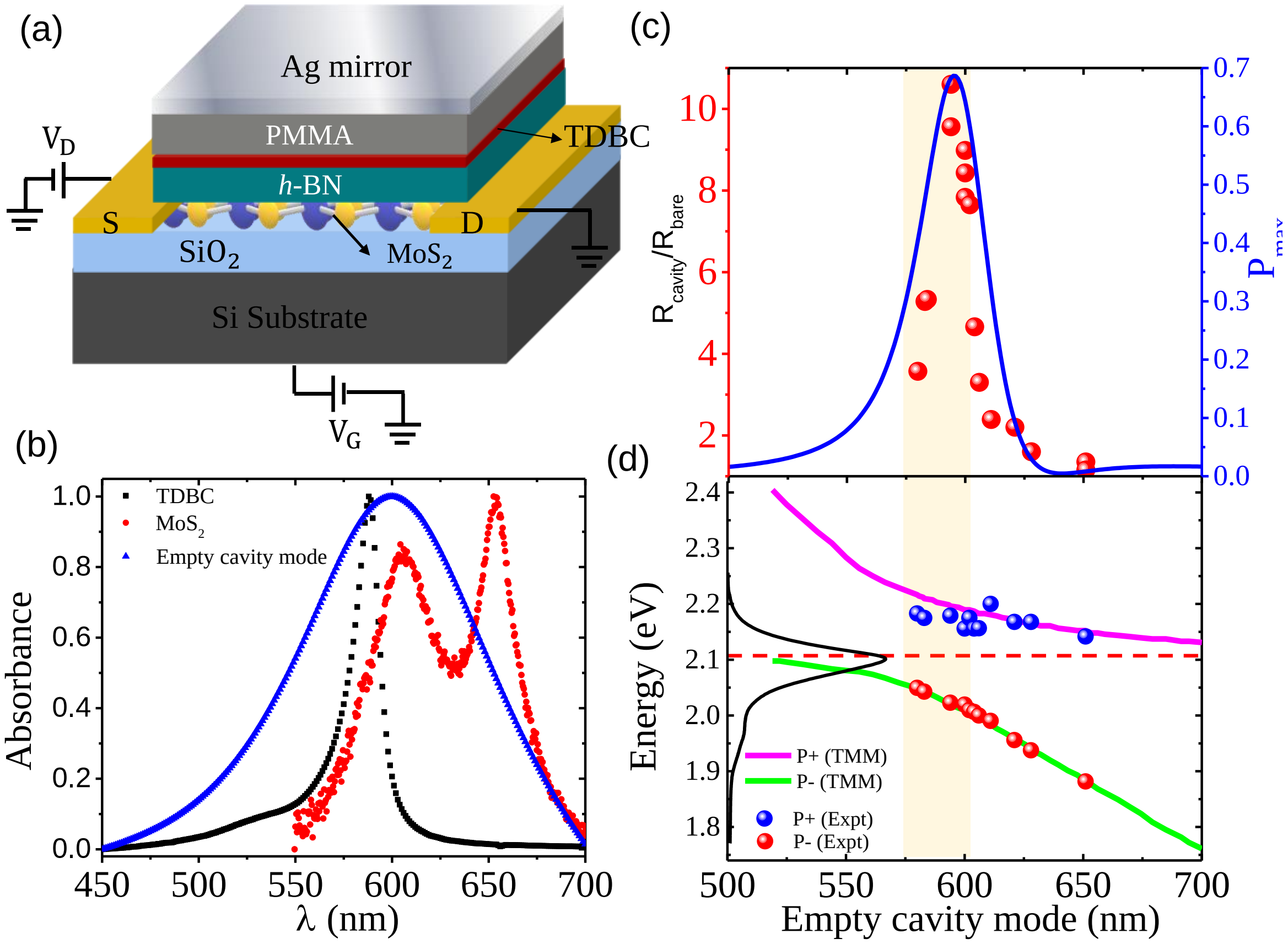

**Figure 4.** *Cavity- electrical output of a quantum phototransistor:* (a) A MOSFET-FP cavity containing a D-A pair placed at a distance of 1500 Å using *h*-BN spacer layer. (b) Normalized absorbance as a function of wavelength for TDBC, $MoS_2$ monolayer and cavity mode are represented by black, red and blue solid lines, respectively. (c) Relative photoresponsivity ($R_{cavity}/R_{bare}$) calculated at 590 nm laser excitation wavelength plotted against the empty cavity mode position (red solid spheres). The blue solid line depicts the quantum mechanically calculated probability of ET (*vide*: Section 13). (d) The increase in photoresponsivity observed in Figure 4c corresponds to the point where the donor absorption intersects with the FP cavity mode. TMM calculated P+ and P- polaritonic states (pink and green solid lines); blue and red solid spheres are corresponding experimental data. The black curve represent the absorbance of the bare TDBC thin film.

To answer this question, we mapped the change in the photoconductivity between the cavity and non-cavity samples. The total photocurrent ($I_{ph}$) of the device is a combination of a photoconductive effect ($I_{PC}$) due to photo-generated charge carriers and a photovoltaic effect ($I_{PV}$).[46] $I_{PV}$ is purely due to the charging of the dielectric layer ($SiO_2$, *h*-BN, TDBC, etc.) that scales with $\Delta V_T$ in the system. For example, adding additional layers such as *h*-BN and TDBC shifts the $\Delta V_T$ to approximately 5V in the dark measurement and an additional -5V upon illumination (Figure S3d). Further, bringing an Ag mirror on the above structure doesn't shift the threshold voltage (Section S10; Figure S9). This indicates that the measured enhancement in photoconductivity is purely due to an increase in the amount of photo-generated charge carriers that occurs under strong coupling. This, along with no change in the electron mobility in the dark and in photo-illuminated samples, indicates that the electronic structure of the $MoS_2$ photoconductor is intact during the experiments. Furthermore, the photocurrent in a cavity confined $MoS_2$/ *h*-BN/TDBC system is conducted with respect to laser excitation power, illustrating a linear response as depicted in Figure S10.

The most salient feature of our experimental observations is that resonant cavity coupling in MOSFET-FP devices enhances their performance due to an increase (~7 times) in the number of photo-generated charge carriers. We now discuss the origin of this observations. The shape of the photoresponsivity spectrum of **Figure 4c** can be approximately reproduced by multiplying the spectral DA overlap function with the relevant polaritonic Hopfield coefficients (details in section S13). This implies that the hybrid D-A nature of polariton states seems to be responsible for the observed device performance. To account for this empirical observation *ab initio* and gain further insight into the operation of polaritonic MOSFET-FP devices, we considered the time evolution of system using the time-dependent Schrödinger equation (ignoring the effects of dissipation) for a system of three coupled resonators: a cavity mode with energy $E_c$, the electronic excited state of a donor species with energy $E_D$ and that of an acceptor with energy $E_A$. Since donors and acceptors are placed at a separation distance larger than the Förster radius, we ignore possible direct donor–acceptor energy transfer. We describe the stationary states of this three–component system with superposition (polaritonic) states:

$$|\Psi\rangle = \alpha\left|D, A; 1_{\mathbf{k},p}\right\rangle + \beta|D^*, A; 0\rangle + \gamma|D, A^*; 0\rangle,$$

where the basis set describes states of the non–interacting system where the energy is in a cavity photon $\left|D, A; 1_{\mathbf{k},p}\right\rangle$ with wavevector $k$ and polarisation $p$, the electronic excited donor state $|D^*, A; 0\rangle$ and the electronic excited acceptor state $|D, A^*; 0\rangle$ respectively. The coefficients $\alpha, \beta, \gamma$ satisfy $|\alpha|^2 + |\beta|^2 + |\gamma|^2 = 1$, and are found as solutions to the equation

$$\hat{H}|\Psi\rangle = E|\Psi\rangle,$$

which also results in a set of three allowed energy values assigned to the lower ($E_L$), middle ($E_M$) and upper ($E_U$) polariton states, and three polaritonic states that we denote as $|LP\rangle$, $|MP\rangle$ and $|UP\rangle$.

As the system is interrogated by shining light onto it, we assume that at time $t = 0$ the state of the system is

$$|\Psi(0)\rangle = \left|D, A; 1_{\mathbf{k},p}\right\rangle,$$

which consists of both D-A species in their electronic ground states and one photon in a cavity mode. Temporal evolution of the system leads to a *redistribution of energy* and at some time $t$, the state of the system is described by

$$|\Psi(t)\rangle = e^{-i\hat{H}t/\hbar}|\Psi(0)\rangle.$$

Following Schäfer *et al*,[9] we quantify "energy transfer" by evaluating the projection of the state of the system at any point through its temporal evolution, on the "transfer target state", namely the one where the energy resides in the electronic excited state of $A$. More concisely, the probability of energy transfer is quantified by evaluating

$$P(t) = |\langle D, A^*; 0|\Psi(t)\rangle|^2$$

a time-dependent function that exhibits (Rabi) oscillations. From this time evolution we obtain the maximum $P_{max}$ and the result of is shown with the blue solid line of **Figure 4c**, which nicely overlaps with the measured photoresponsivity enhancement.

This agreement suggests that the role of strong light-matter coupling in these experiments is to spatially delocalize energy inside the mode volume of the cavity. This delocalization in turn, results in the population of acceptor-like polariton states through ultra-fast Rabi oscillations that take place in the cavity-donor-acceptor system (Figure S12). The population of acceptor-like polaritonic states is ultimately responsible for the enhanced photoresponsivity. Maximum delocalization occurs at the point of cavity-donor resonance, which takes place at energies above the exciton resonances of uncoupled $MoS_2$, consequently resulting in the "sensitization" of the MOSFET-FP device at this incident photon wavelength. Our simple model predicts that further improvements could be obtained by increasing the cavity-donor coupling strengths, which could be achieved by a number of strategies, including increasing the cavity quality factor.

## Conclusion

We have shown that spatially separating a donor-acceptor pair in a cavity result in the formation of entangled polaritonic states. Further, we prepared room temperature quantum phototransistors using single mirror MOSFET configuration, and studied energy transfer to an atomically thin $MoS_2$ monolayer by photoresponsivity. The photoresponsivity of more than a dozen independent phototransistors directly correlates with the spectral signature of the polaritonic states. Both optical and electrical measurements allowed us to approximate the patterns in energy migration and to assess the significance of polaritonic states in facilitating long-range energy transfer. Further we simplified the model using TMM and *ab initio* calculations to explicitly prove the entangled nature of the donor-acceptor polaritonic states. Leveraging the collective behaviour and modification of probability density is an attractive tool that strong coupling can offer to prepare new-generation polaritonic devices. Our system maybe particularly of interest for developments in applications such as in quantum batteries.[47]

## Supporting Information

Detailed studies of optical, electrical, and microscopic characterization of phototransistor devices, TMM simulation, and quantum mechanical calculation of the ET probability is given in the supporting information.

## Acknowledgements

J. D. thank DST (India) for the Women Scientist Scheme-A (WOS-A) Fellowship. J. G. wishes to acknowledge IISER Mohali start-up grant for conceiving the project and the provision of laboratory facilities. P. B. and K. K. thank IISER Mohali for the PhD fellowship. We greatly appreciate the assistance provided by Mr. Devansh Swadia, and department of biological sciences, IISER Mohali for conducting AFM studies.

**Keywords:** Photoresponsivity, polaritonic states, J-aggregates, 2D materials, organic-inorganic hybrid system.

# Supporting Information

## Table of Contents

## Experimental methods

### Sample Fabrication:

In the LBL process, the substrates undergo sequential immersions in cationic and anionic solutions (SICAS) for 5 minutes each, followed by a 1 min rinse in deionized water to build up a thin film. PDAC (poly (diallyl dimethylammonium chloride)) is the cationic solution ($6 \times 10^{-2}$ M) prepared in deionized water and the anionic solution ($1\times10^{-4}$ M) is, J-aggregate cyanine dye TDBC (5,6-dichloro-2-[3-[5,6-dichloro-1-ethyl3-(3-sulfopropyl) 2(3H)-benzimidazolidene]-1-propenyl]-1-ethyl-3-(3-sulfopropyl) benzimidazolium hydroxide, inner salt) in deionized water. Monolayer flakes of $MoS_2$ and large *h*-BN flakes were mechanically exfoliated using the scotch tape method from a synthetic bulk crystal (HQ Graphene, The Netherlands) on thin PDMS stamps, and later on, transferred to a gold electrode. Monolayer $MoS_2$ flake was confirmed by both absorption and photoluminescence measurements. The thickness of the multilayer *h*-BN flake was confirmed by AFM height analysis (tabletop AFM; Bruker).

A n++ doped silicon wafer with 90 nm $SiO_2$ with an electrode pattern of 10/30 nm Ti/Au was purchased from Fraunhofer-IPMS, Germany. Then, using a homebuilt transfer stage, desired flakes were dry transferred utilizing surface adhesion onto the Ti/Au electrode (L = 5 μm). Prior to electrical measurements, samples were annealed at 200 °C for 2 hours in an inert gas glovebox. On a thick PDMS substrate, a Ag mirror was sputtered (40 mA; BT300, Hind high-vacuum) to create the FP cavity in the FET configuration. PMMA (Polymethyl methacrylate, 120 kDa; Sigma-Aldrich) was spin-coated (LABSPIN 6, Suss-Microtech) on an Ag mirror to obtain various thicknesses. Here we used a 2-4% PMMA solution in Anisole and spin-coated at different rpm (with a difference of 100 rpm) on a Ag-coated PDMS substrate to control the resonance of the cavity. Specifically, the ON resonance sample is achieved by spin coating 3% solution at 2400 rpm, with an acceleration of 4000 rpm/s. Finally, the PMMA-coated Ag mirror was diligently positioned onto the exposed FET device housing an active $MoS_2$ layer.

## Section 1. Absorption and PL spectra of donor TDBC and acceptor $MoS_2$

Normalized absorption and PL spectra of 10 nm TDBC thin film coated on optically clean glass substrates show peaks at 590 nm and 594 nm, respectively (**Figure S1 (a)).** The FWHM of TDBC is 17 nm and the estimated Stoke shift is only 4 nm. **Figure S1 (b)** displays the normalized absorption and PL of the $MoS_2$ monolayer exfoliated on the PDMS substrate, where absorption spectra show two exciton peaks at 606 nm (B-exciton) and 655 nm (A-exciton) and a PL peak around 655 nm.

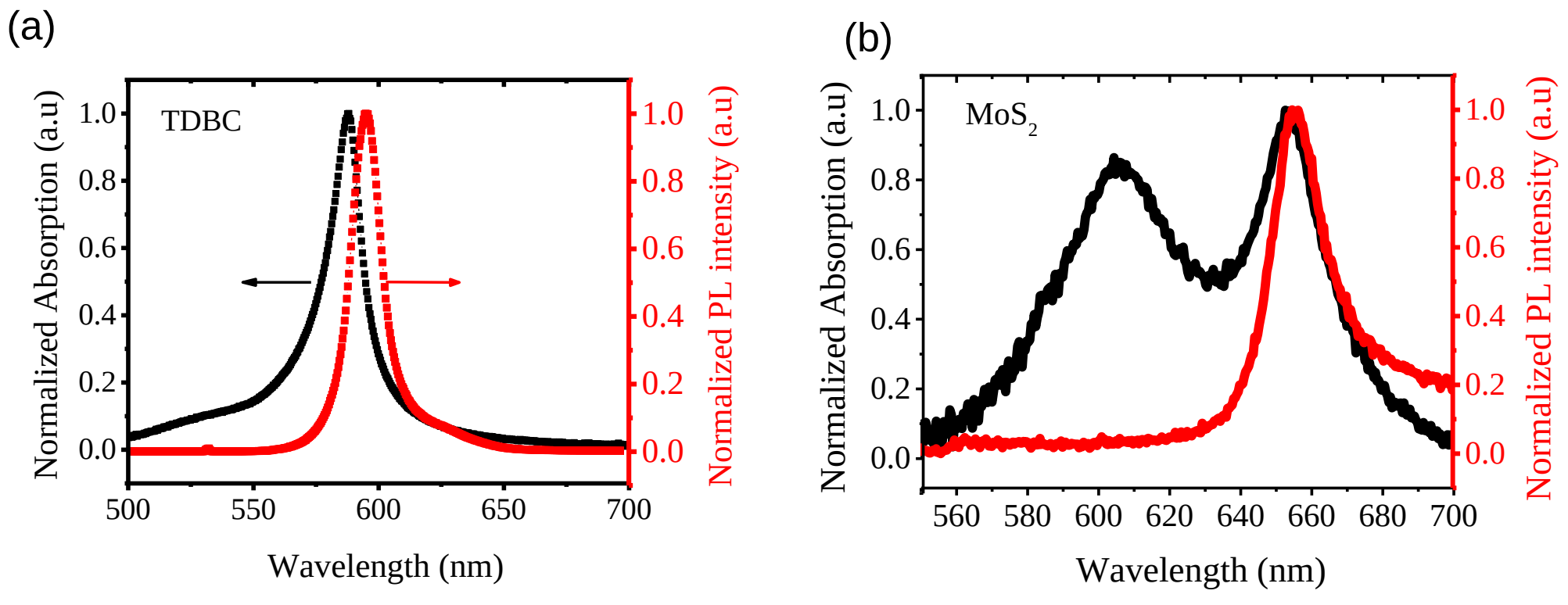

Figure S1: Absorption and emission of (a) TDBC J-aggregates and (b) $MoS_2$ monolayer

## Section 2. Refractive indices of TDBC and $MoS_2$ monolayer

Figure S2 (a) and (b) show the refractive indices of the TDBC and $MoS_2$ monolayers respectively, extracted from the absorption spectra in Figure S1 (a) and (b). The imaginary part of the refractive indices emphasizes the sharp absorption characteristics of the TDBC compared to monolayer $MoS_2$ exciton band.

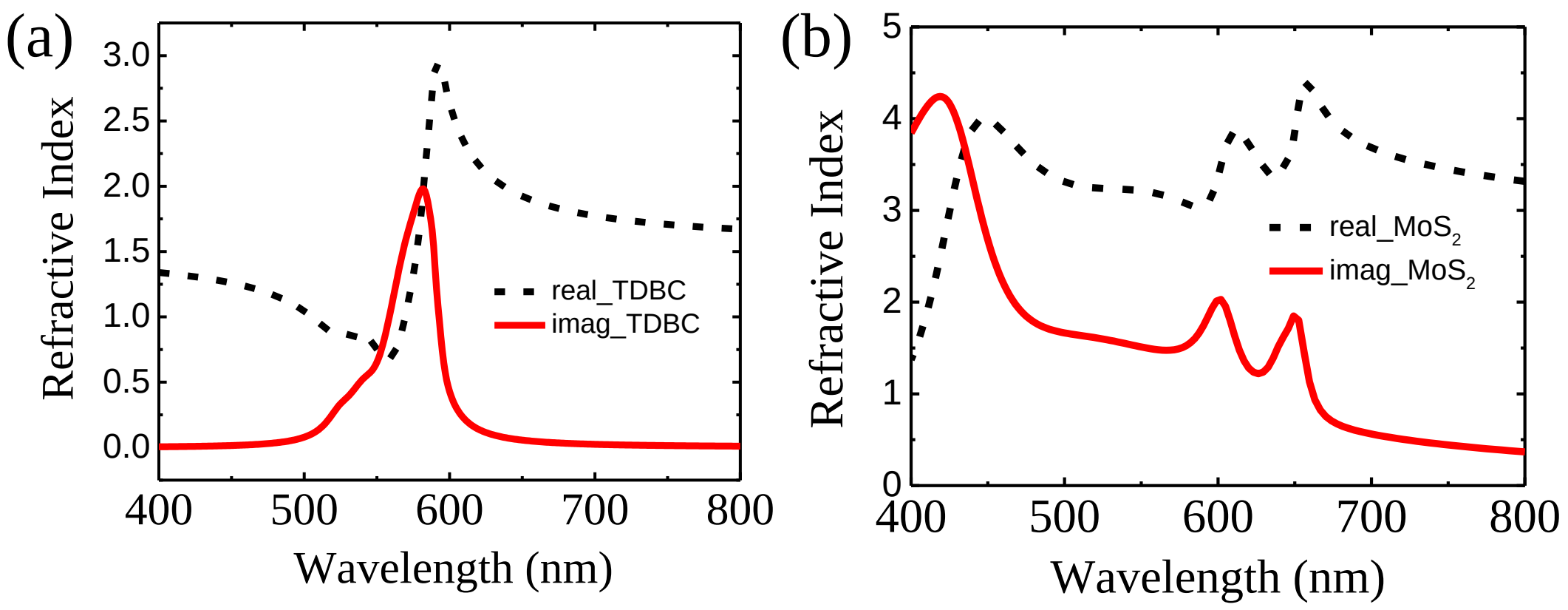

**Figure S2:** The real and imaginary refractive indices of (a) 10 nm LBL TDBC and (b) $MoS_2$ monolayer calculated from the absorption spectra (extracted from Figure S1).

## Section 3. Photocurrent and photoresponsivity spectra as a function of laser excitation wavelength for non-cavity system with *h*-BN spacer layer

Figure S3(a) displays the photocurrent spectra for $MoS_2$/*h*-BN and $MoS_2$/*h*-BN/TDBC, which get affected if the h-BN layer is thin (within the Förster zone; 2 nm), and the corresponding photoresponsivity is shown in Figure S3(b).

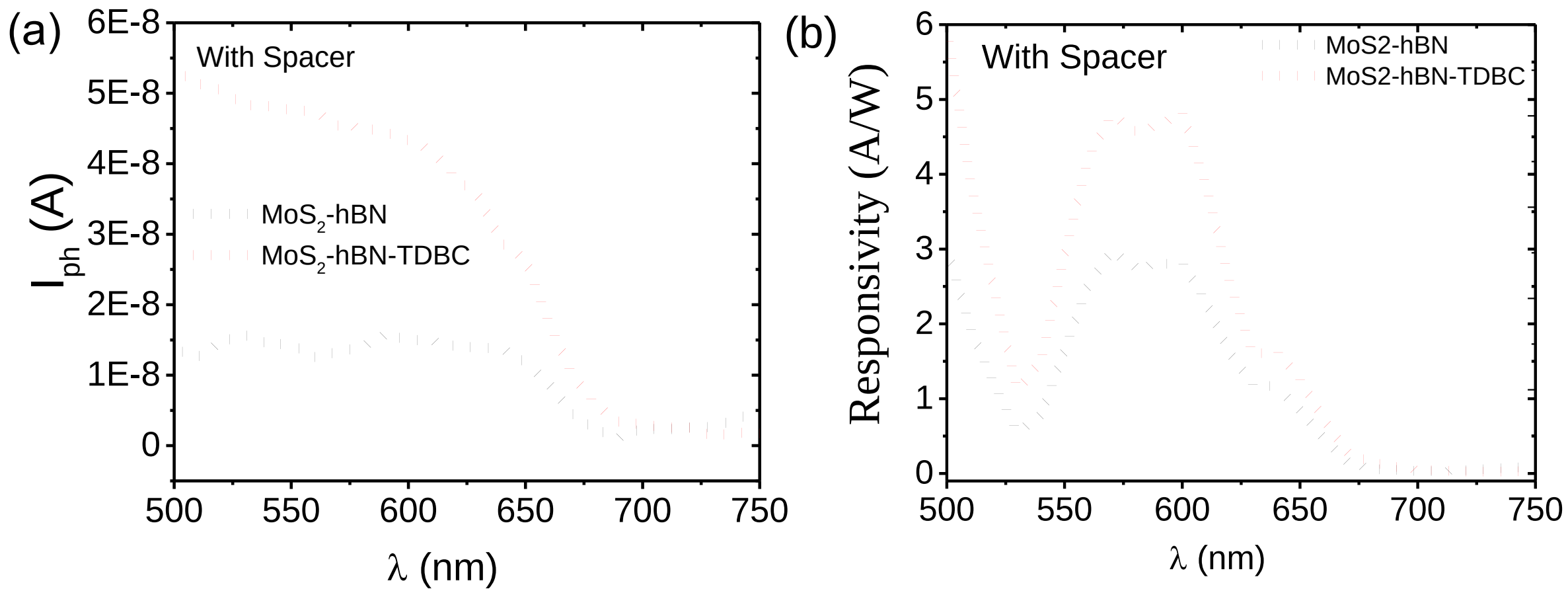

Figure S3: (a) Photocurrent spectra of $MoS_2$ monolayer, before (black) and after (red) TDBC deposition with *h*-BN spacer layer (fixed spacer thickness of 2 nm). (b) The photoresponsivity spectra of MoS2/h-BN and MoS2/h-BN/TDBC composite systems corresponding to the photocurrent are shown in (a).

## **Section 4: FRET efficiency calculation**

$MoS_2$ crystal length (a) = 0.316 nm,

Monolayer height (L) = 0.8 nm.

$MoS_2$ absorption, A= ε*L*C

where, ε= molar absorption, C= number of Mo-S bonds per unit volume.

For the $MoS_2$ monolayer, one Mo atom is surrounded by 6 'S' atoms.

So, $C = \frac{6}{N_A * \text{Volume}}$, Hexagonal lattice, Volume $= \frac{\sqrt{3}}{2} a^2 c$, $N_A$ = Avogadro number.

Forster Energy Transfer Efficiency,[48]

$\eta_{FRET} = \frac{1}{1+(r/R_0)^6}$, Where $R_0$ = Förster radius,

$R_0^{\ 6} = \frac{9 \ln (10)}{128\Pi^5 N_A} \frac{k^2 \phi_D}{n^4} J$; k= orientation factor = $\frac{2}{3}$, $\phi_D$ = quantum yield of TDBC = 0.02,

$n = \frac{n_{TDBC} * L_{TDBC} + n_{MoS_2} * L_{MoS_2}}{L_{TDBC} + L_{MoS_2}}$;

$L_{TDBC}$=10 nm, $L_{MoS2}$ = 0.8 nm

$n_{TDBC}$ (at TDBC PL wavelength), $n_{MoS2}$ (at wavelength corresponds to$MoS_2$ absorbance )

Overlap Integral, J = $\frac{\int f_D(\lambda)\, \varepsilon_A(\lambda) \lambda^4 d\lambda}{\int f_D(\lambda) d\lambda}$, where

$f_D(\lambda)$ is the donor emission spectrum normalized to it's area

$\varepsilon_A(\lambda)$ is the molar extinction coefficient of acceptor.

The FRET efficiency ($\eta_{FRET}$) can be calculated from the photoresponsivity enhancement by,

$$\eta_{FRET} = \frac{E_{FRET}}{6.3 * \phi_D * k^2}$$

$E_{FRET} = \frac{\Delta R_{ET}}{R_{MoS_2}}$; ratio of change in photoresponsivity after depositing TDBC to the bare $MoS_2$.

6.3 is the absorption ratio of J-aggregate to $MoS_2$ at the maximum responsivity, which occurs when the J-aggregate film exhibits its maximum absorption.

**<u>Section 5: Atomic force microscopy (AFM) images of *h*-BN spacer layer</u>**

AFM images of thick *h*-BN spacer layers are shown in Figure S4. These layers have an approximate thickness of about 150 nm, which we utilized in cavity devices, and approximately 8 nm, employed in non-cavity devices.

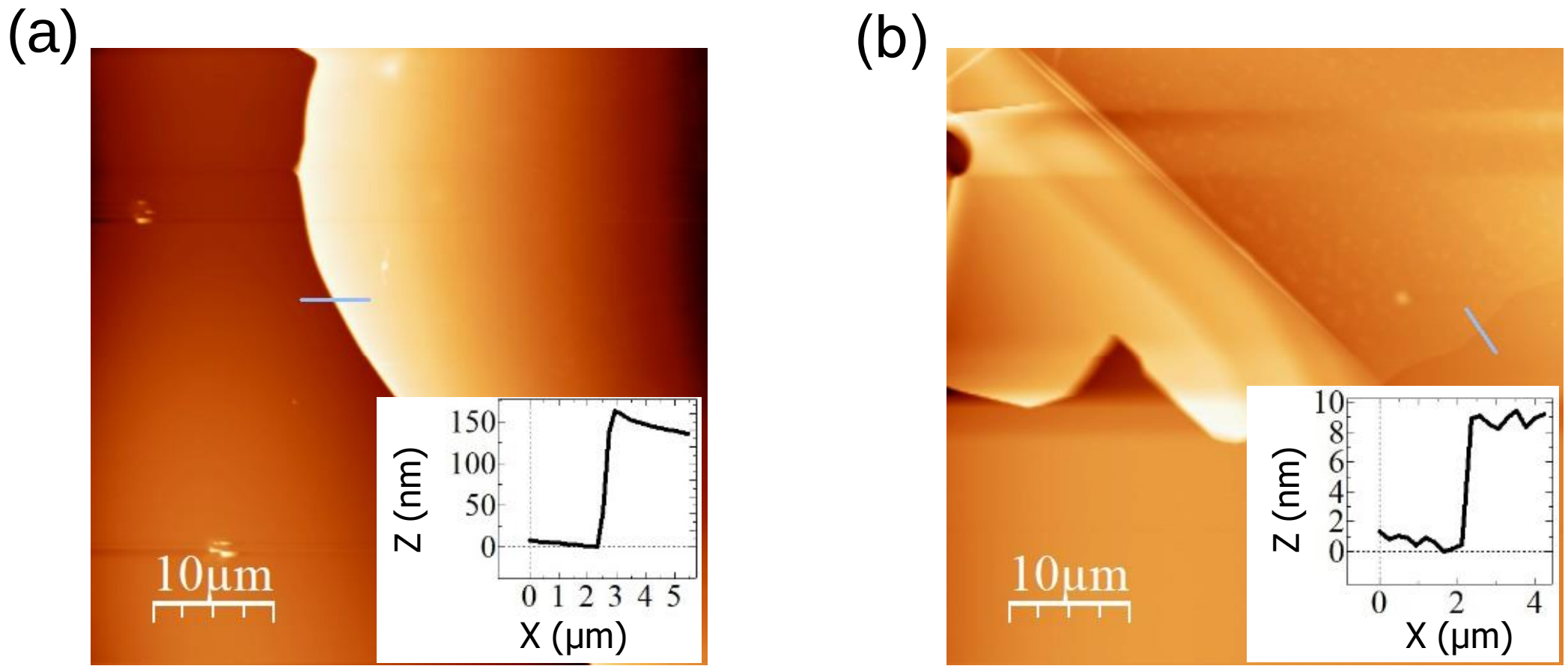

**Figure S4:** AFM images of the *h*-BN multilayers having approximate thicknesses of (a) 150 nm and (b) 8 nm.

## Section 6. Photoconductivity ON/OFF ratio for $MoS_2$ monolayer

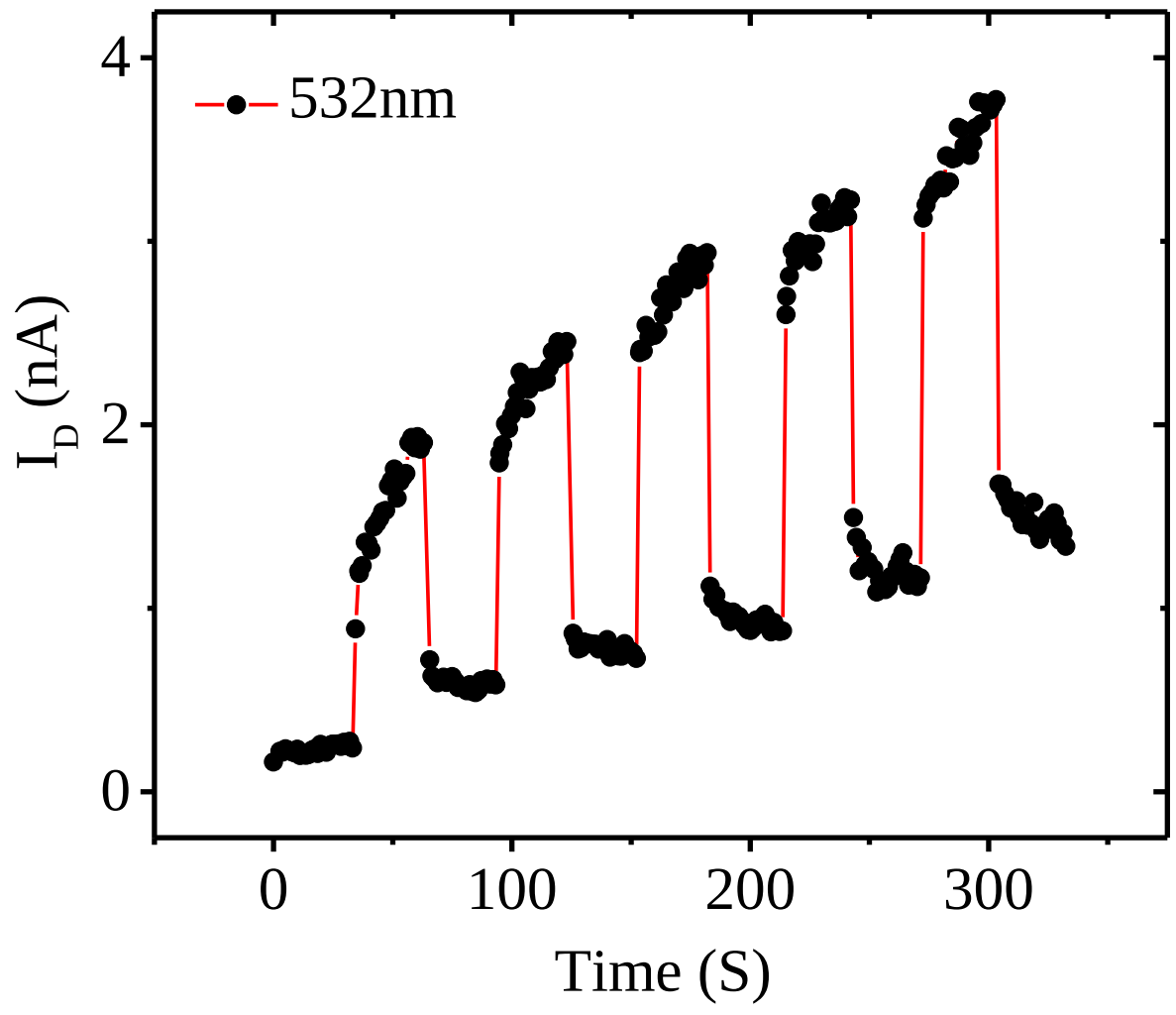

**Figure S5:** A representative ON/OFF photocurrent cycle of bare $MoS_2$ monolayer MOSFET device.

ON/OFF photocurrent cycle of $MoS_2$ monolayer FET device for laser illumination at 532 nm are shown in Figure S5. Here, drain voltage and gate voltage are fixed at 0.1 V and -8 V, respectively. We fabricated various devices with distinct geometrical shapes through a mechanical exfoliation process for our tuning experiments. It's important to note that Schottky barrier height at the Au /$MoS_2$ interface can vary due to these different geometrical shapes. To diminish the imapct of Schottky barrier height fluctuations, we employed the relative change in photoresponsivity as a normalization technique, allowing us to isolate and analyze the influence of strong coupling effects.

## Section 7. Photoresponsivity spectra of $MoS_2$/*h*-BN (*only*) inside an FP cavity

As a clean control experiment, we probed the photoresponsivity spectra. $MoS_2$/*h*-BN alone was taken inside the cavity by tuning it to ON-resonance condition. photoresponsivity spectra show negligible variation for bare $MoS_2$/*h*-BN and $MoS_2$/*h*-BN-cavity devices. All the photoresponsivity data is calculated using $V_d$ = 0.5 V.

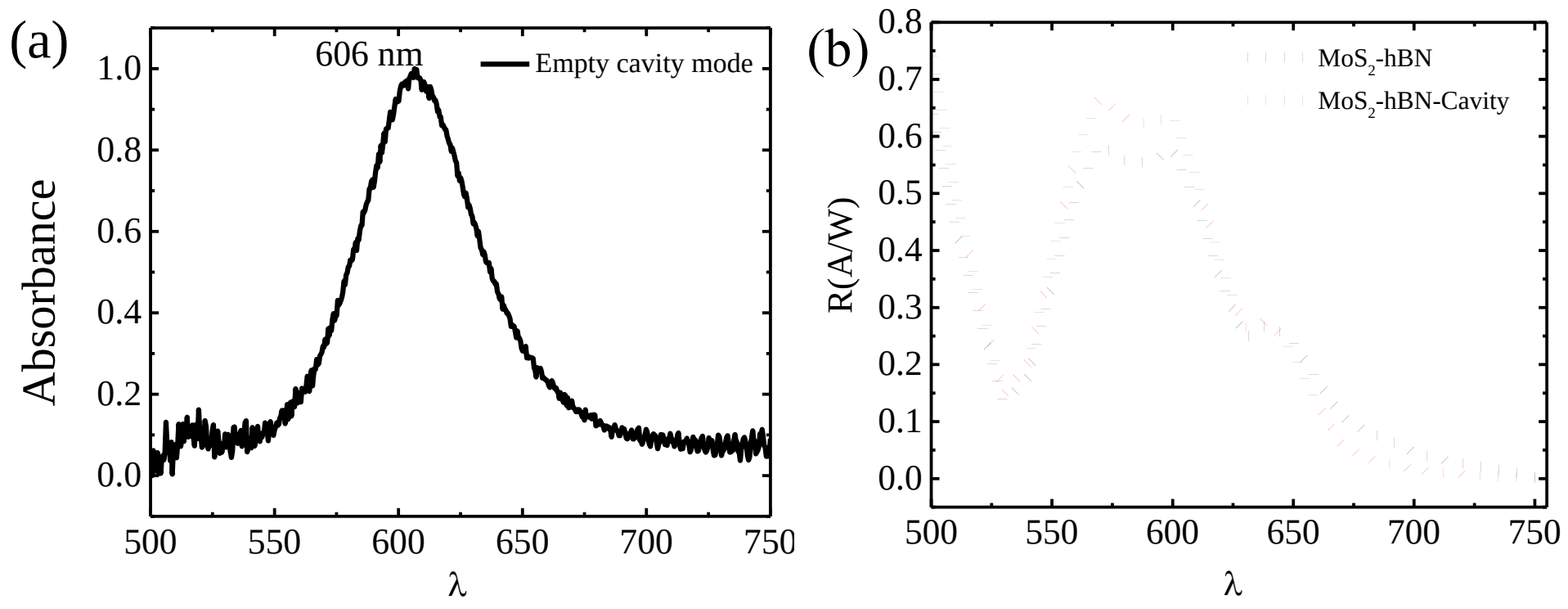

**Figure S6:** (a) Optical spectra (1-R) of $MoS_2$/*h*-BN *only* inside the MOSFET cavity. (b) Photoresponsivity spectra of bare $MoS_2$/*h*-BN, and $MoS_2$/*h*-BN ON resonance cavity devices.

**Section 8: Reflection spectrum and photoresponsivity as a function of laser excitation wavelength for cavity: 594 nm cavity (ON-resonance; raw data of Figure 4c)**

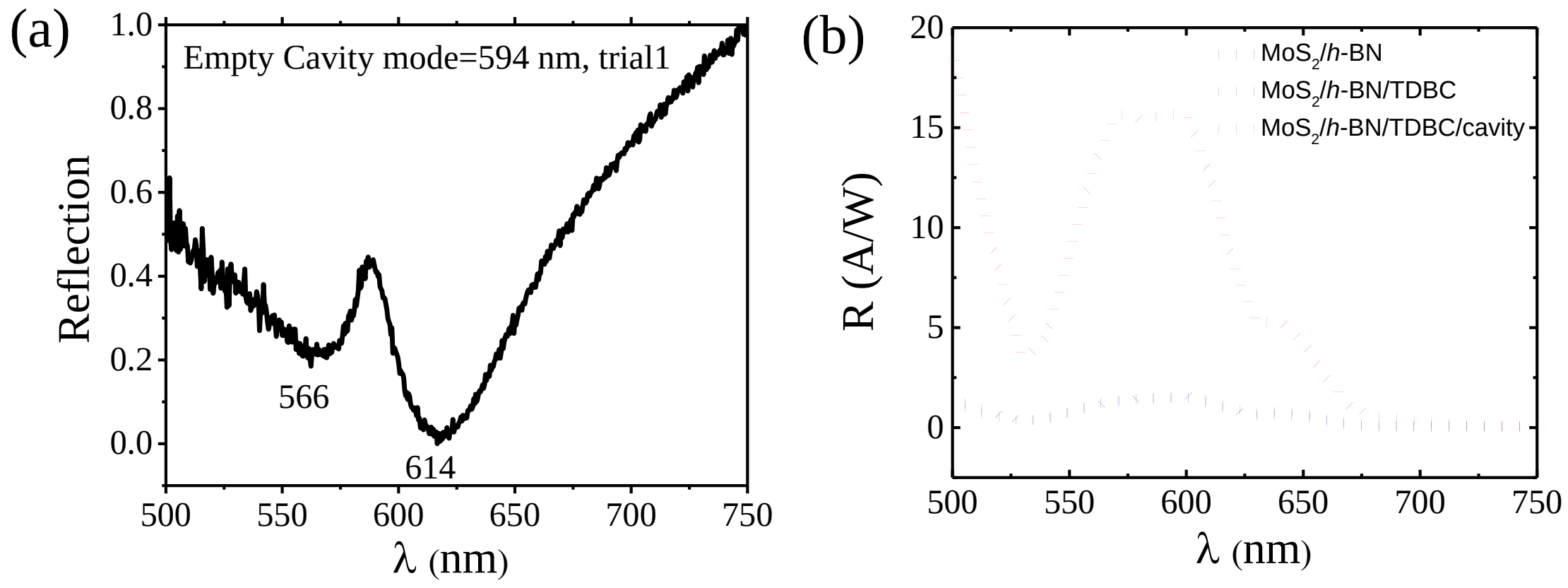

**Figure S7:** (a) Normalized cavity reflection spectra measured at ON-resonance as a function of wavelength showing the formation of polaritonic states p+ (566 nm) and p- (614 nm). These states correspond to the calculated empty cavity mode at 594 nm, determined via TMM. (b) Photoresponsivity spectra for $MoS_2$/*h*-BN (black), $MoS_2$/*h*-BN/TDBC (red), and $MoS_2$/*h*-BN/TDBC/cavity (blue; ON-resonance cavity; *h*-BN = 150 nm).

**Section 9. Raw data of the optical (left side) and electrical (right side) output of different cavity modes as measured in Figure 4c.**

MOSFET-FP cavities were fabricated by varying the thickness of the PMMA spacer while keeping the h-BN and TDBC thicknesses fixed. Different PMMA thickness values resulted in varying positions of the cavity modes.

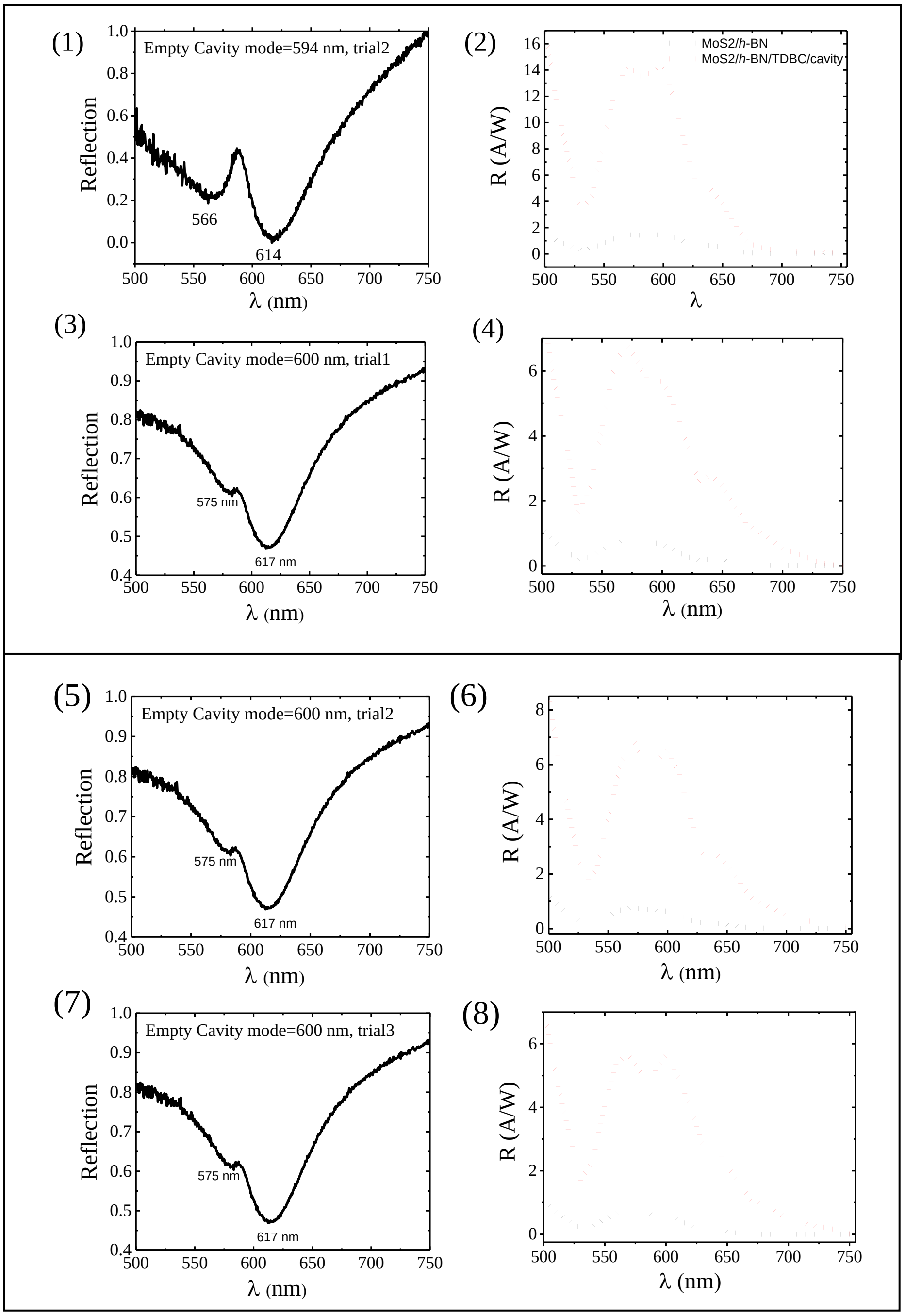

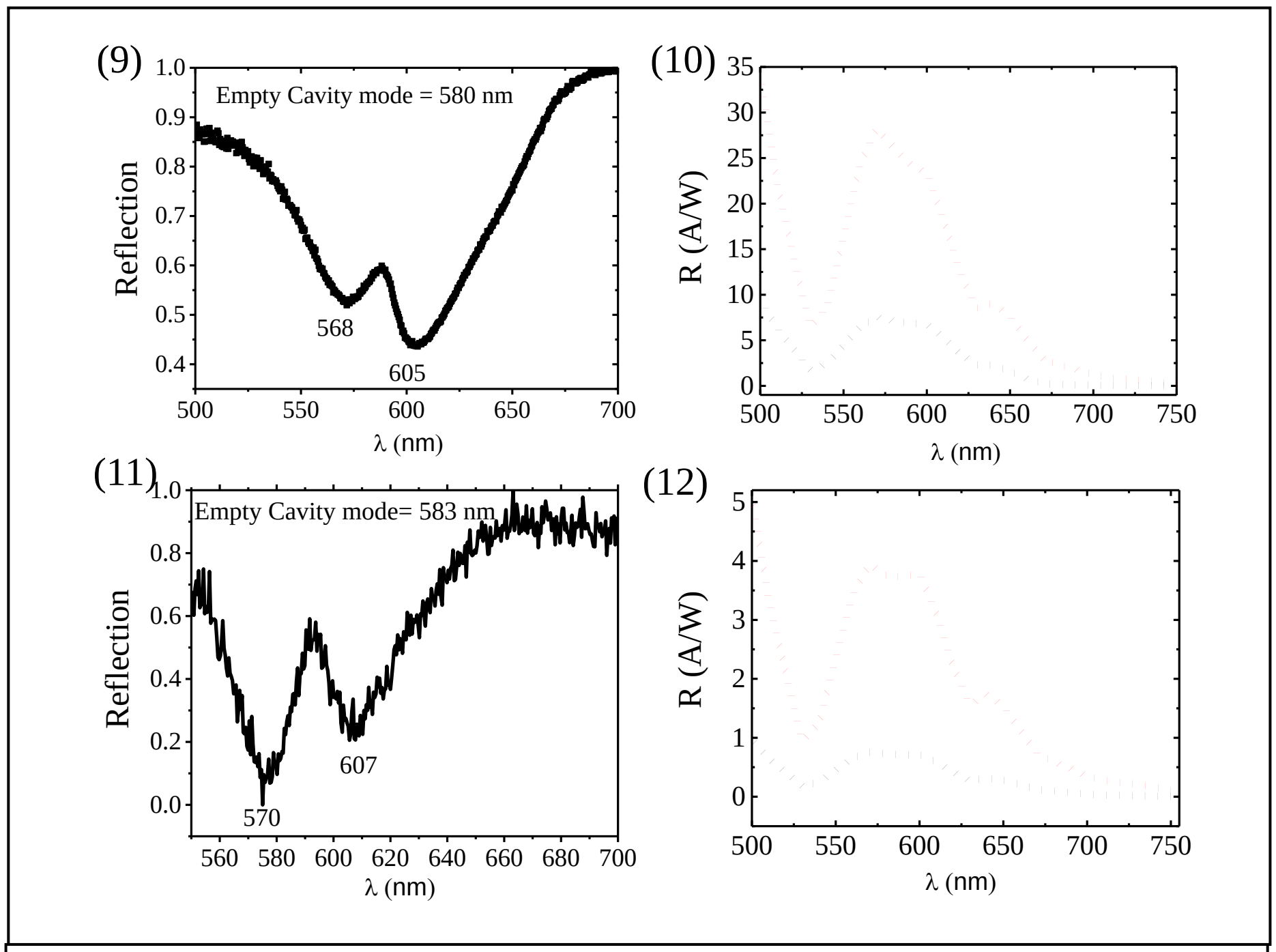

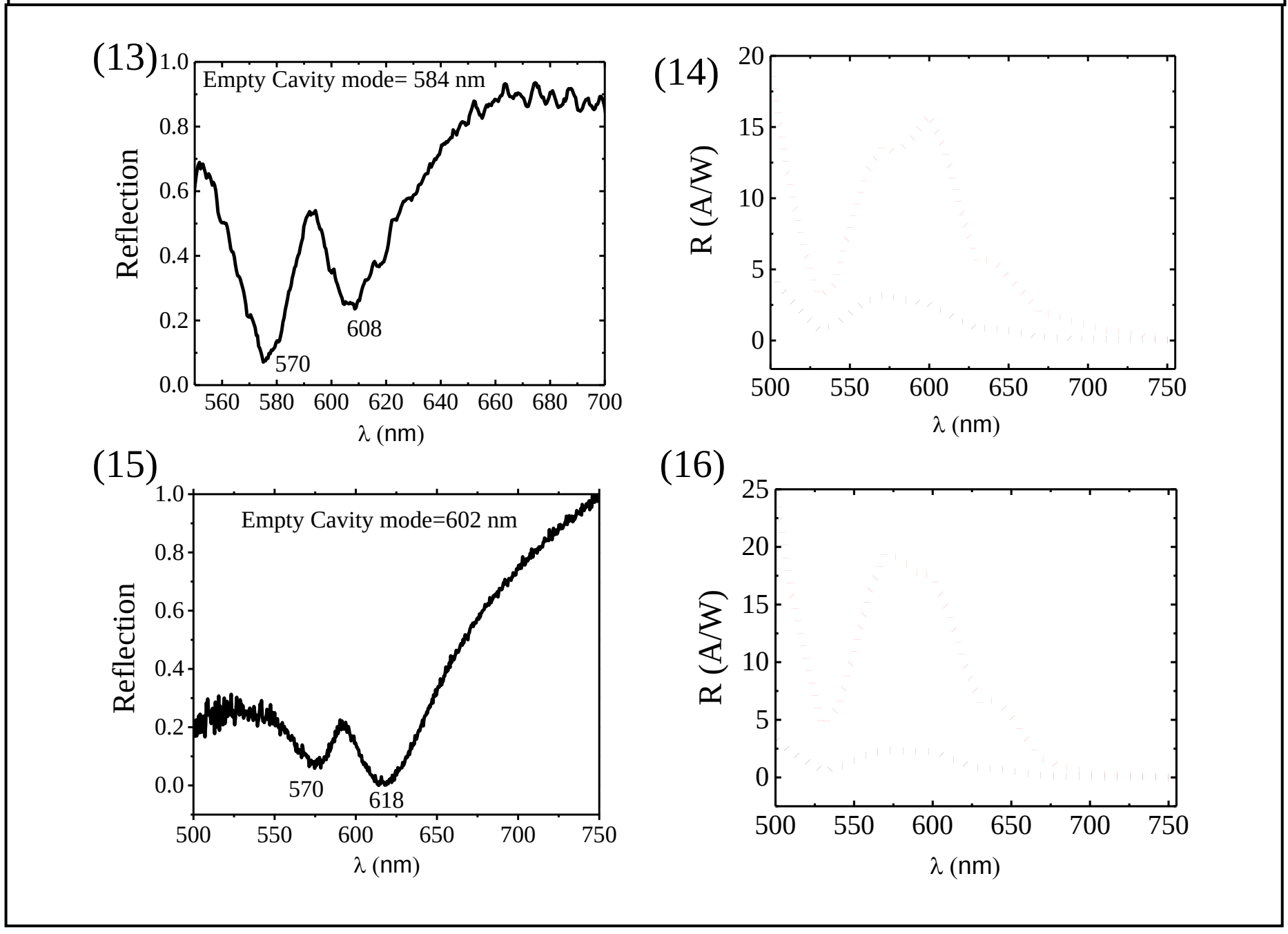

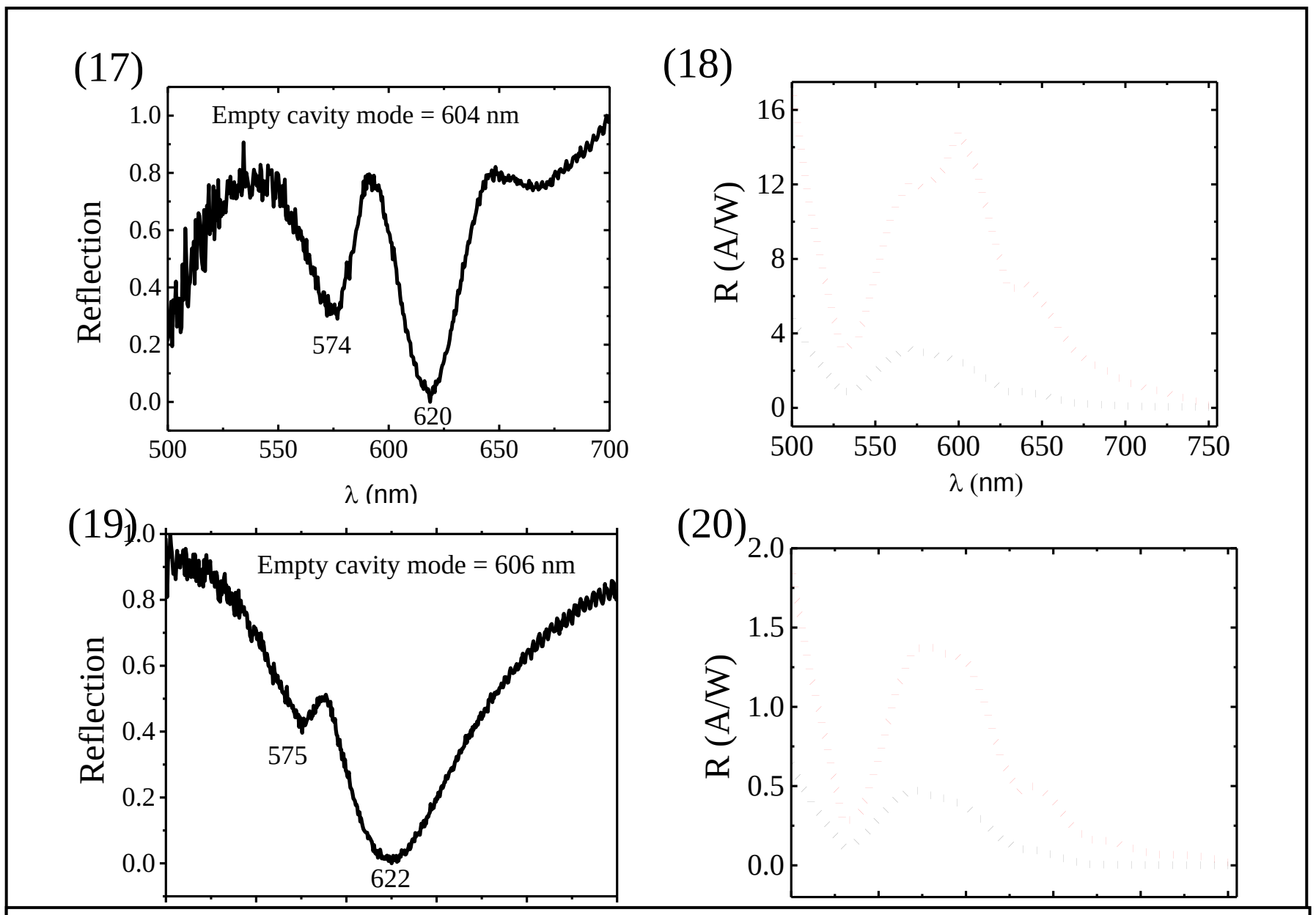

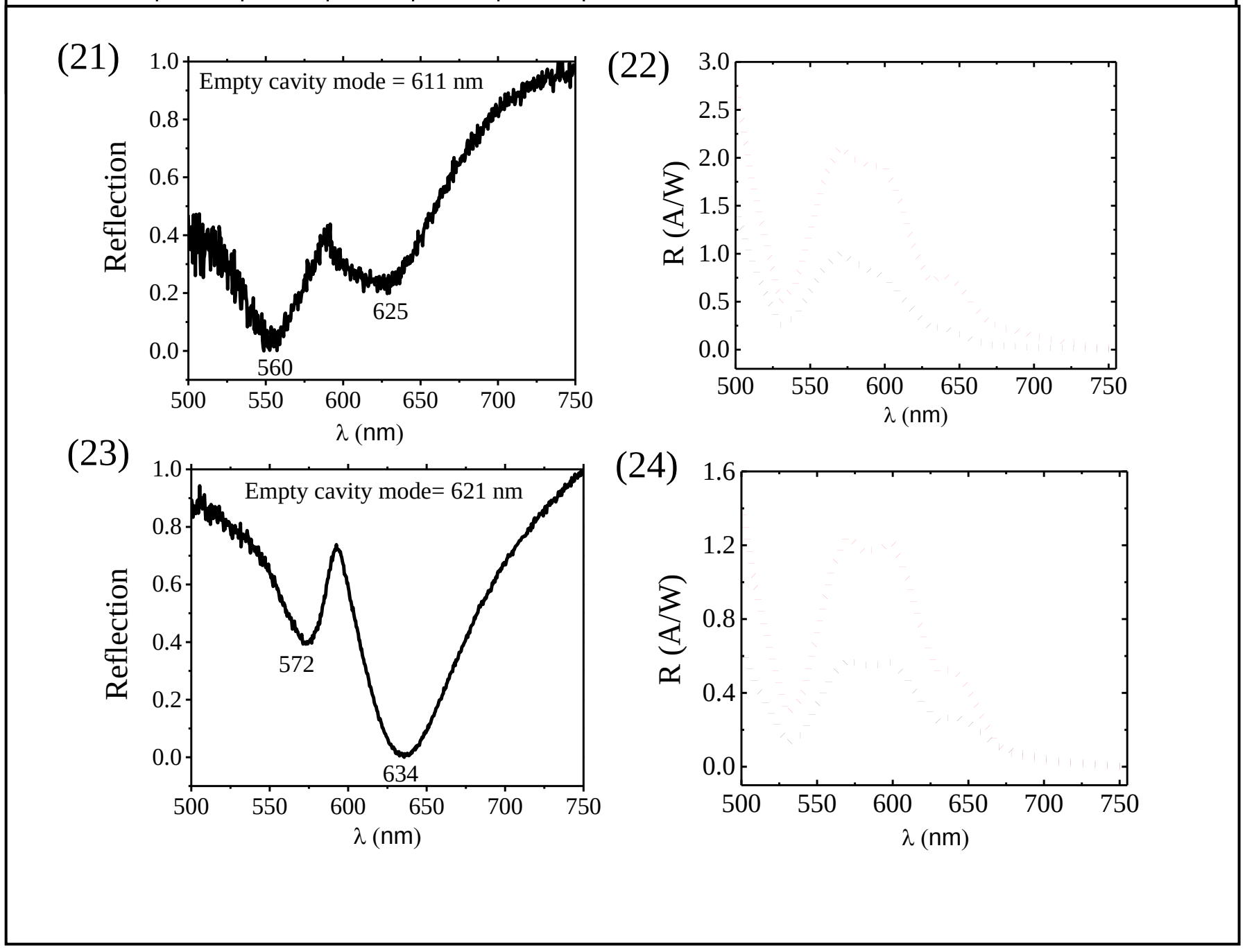

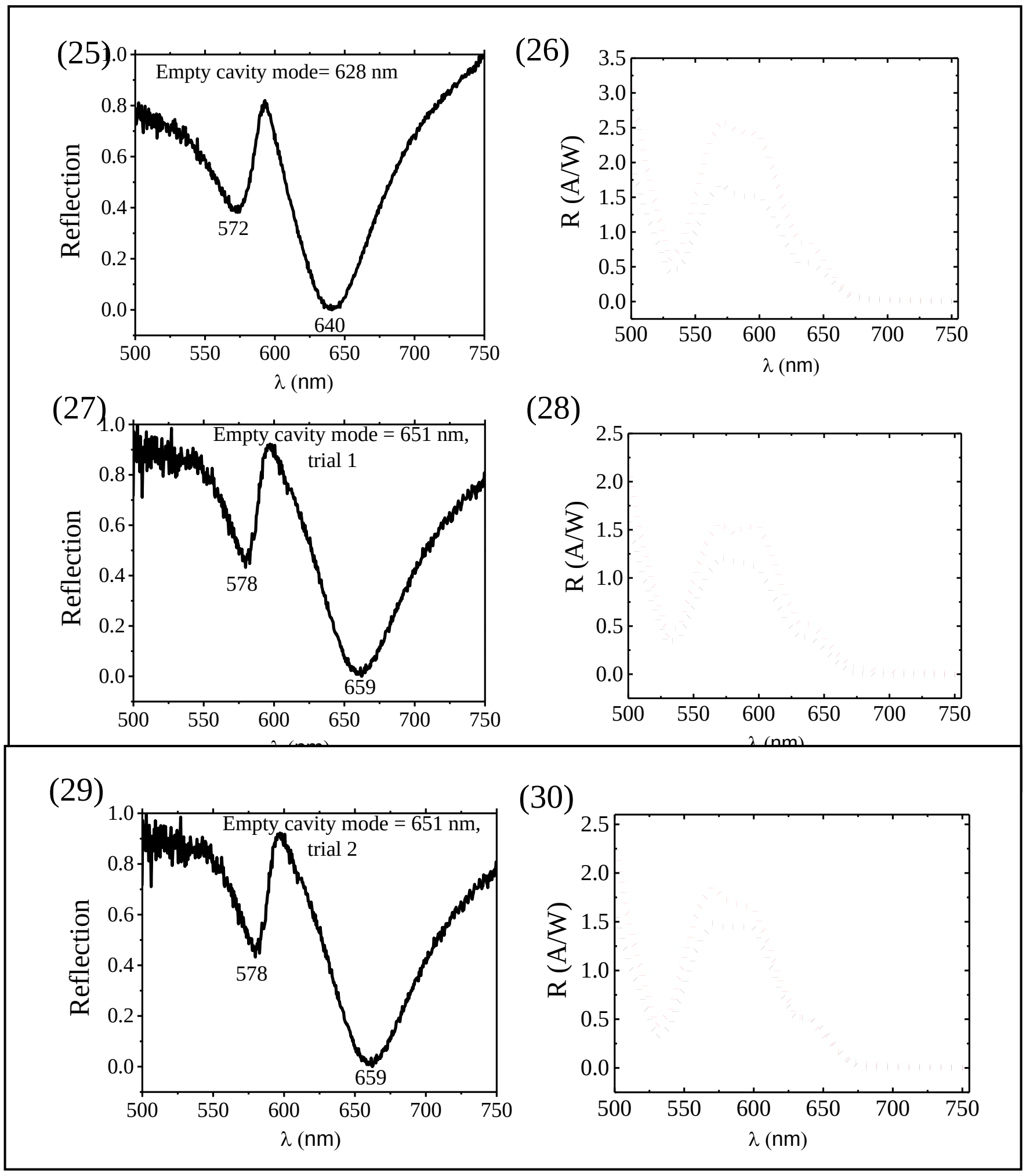

**Figure S8:** Raw data of the optical (left side) and electrical (right side) output of different cavity modes as measured in Figure 4c.

## Section 10. $I_D$-$V_G$ plot for ON-resonance cavity

$I_D$-$V_G$ characteristics of bare $MoS_2$/*h*-BN, $MoS_2$/*h*-BN/TDBC and $MoS_2$/*h*-BN/TDBC/cavity are shown in Figure S9. Threshold voltage of $MoS_2$/*h*-BN/TDBC shifted from -10V to -15V. Further, bringing Ag mirror on the device doesn't shift the threshold voltage.

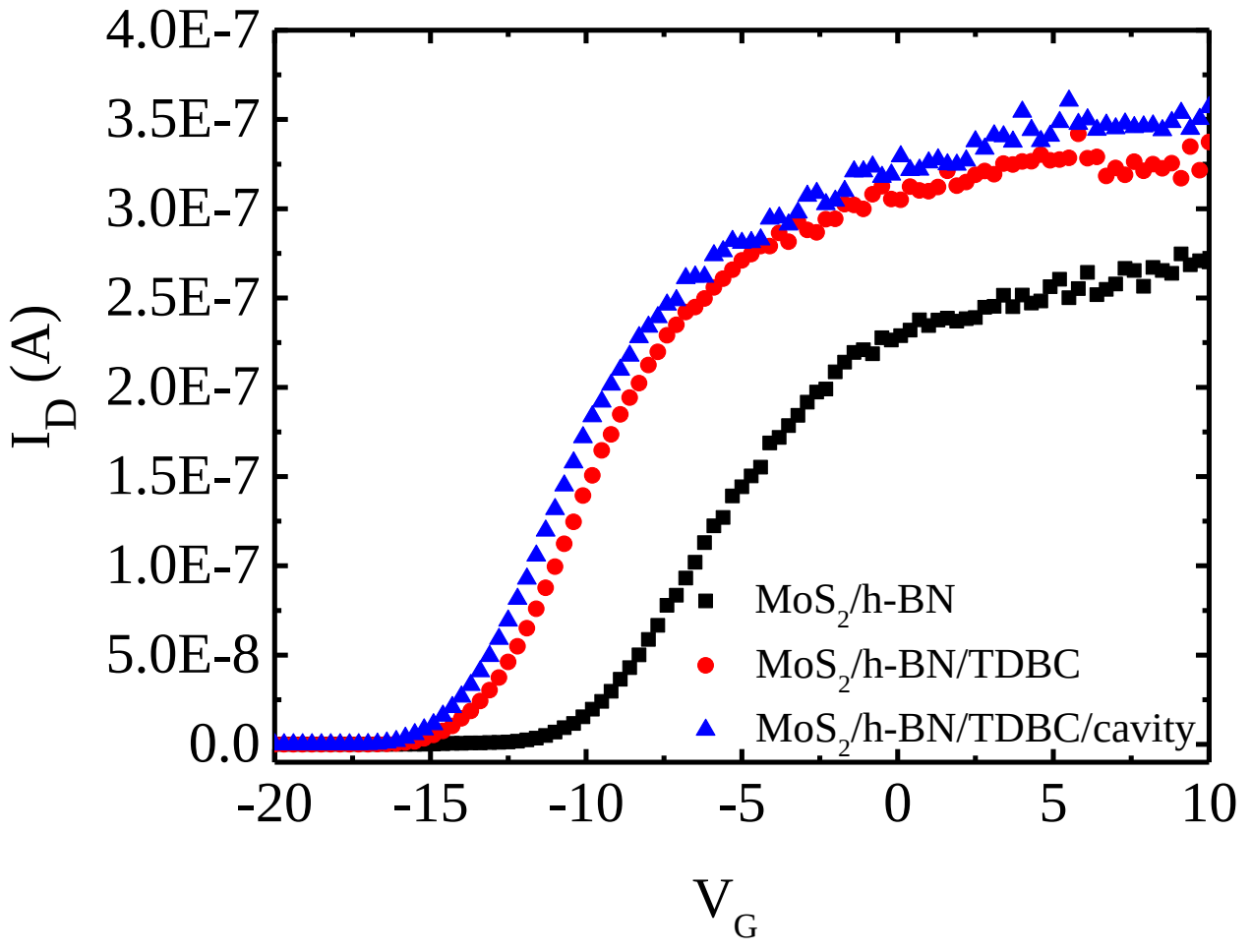

**Figure S9:** $I_D$-$V_G$ characteristics of bare $MoS_2$/*h*-BN, $MoS_2$/*h*-BN/TDBC and $MoS_2$/*h*-BN/TDBC/cavity at an excitation wavelength of 590 nm, where the source - drain bias voltage is at 0.5 V.

## Section 11. Photocurrent as a function of laser excitation power

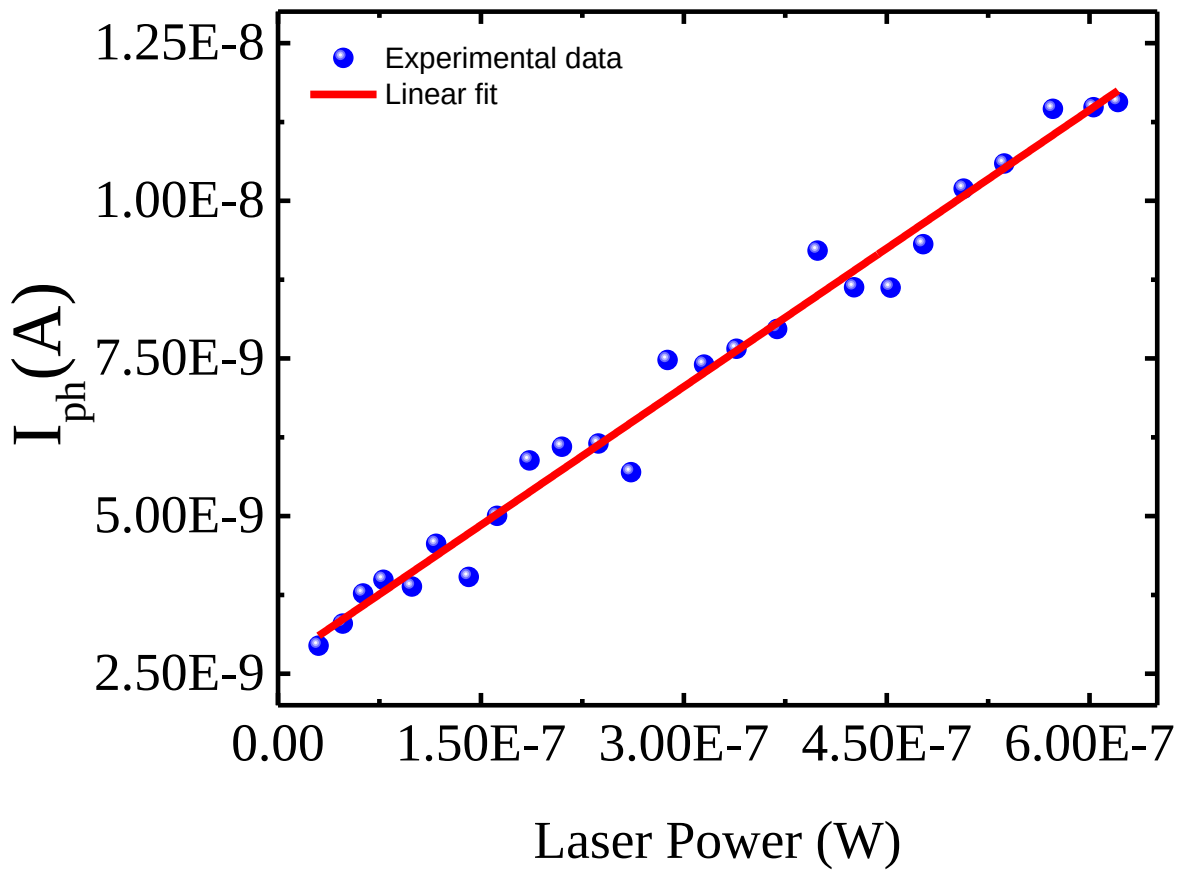

**Figure S10:** Photocurrent in a cavity confined $MoS_2$/h-BN/TDBC system as a function of laser excitation power.

## Section 12. Number of photogenerated charge carriers

Total photocurrent $I_{ph}= I_{PC} + I_{PV}$, [46] (1)

The photovoltaic (PV) effect, also known as the "photogating effect", can be defined as the alteration of a transistor's threshold voltage$V_T$ when transitioning from a dark state to an illuminated state, resulting in a change to $V_T$ - $\Delta V_T$.

$I_{PV} \approx g_m \Delta V_T$ (2)

Where $g_m = dI_D/dV_G$ is the transconductance.

The photoconductive (PC) component of the photoresponse corresponds to the rise in conductivity, denoted as $\Delta\sigma$, resulting from the presence of excess carriers generated by illumination.

$I_{PC} = (W/L)V_D \Delta\sigma$ (3)

$$\frac{(I_{PC})\text{cavity}}{(I_{PC})\text{bare}} = \frac{(I_{ph}-I_{PV})\text{cavity}}{(I_{ph}-I_{PV})\text{bare}} \quad (4)$$

Since $\sigma = ne\mu$ (5),

where n=number of charge carrier, e = electronic charge, and $\mu$=mobility (calculated from $I_D$-$V_G$ plot).

Here, $n_{cavity}/n_{bare}$ = 7.45 (~7 times at ON-resonance)

**Section 13: Hopfield Coefficient:**

The Hopfield coefficients are determined by fitting the dispersion curve obtained from solving a 3x3 matrix of coupled oscillator model.

$$\begin{bmatrix} \mathbf{E_c - i\gamma_c} & \mathbf{g_A} & \mathbf{g_D} \\ \mathbf{g_A}^* & \mathbf{E_A - i\gamma_A} & \mathbf{0} \\ \mathbf{g_D}^* & \mathbf{0} & \mathbf{E_D - i\gamma_D} \end{bmatrix} \begin{bmatrix} \alpha \\ \beta \\ \gamma \end{bmatrix} = \mathbf{E_{pol}} \begin{bmatrix} \alpha \\ \beta \\ \gamma \end{bmatrix}$$

Where $E_C$ is the empty cavity mode positions for various cavities. $E_A$ and $E_D$ are the exciton position for acceptor ($MoS_2$) and donor (TDBC). In our case, $E_A$ = 2.046 eV; $E_D$ =2.1 eV;

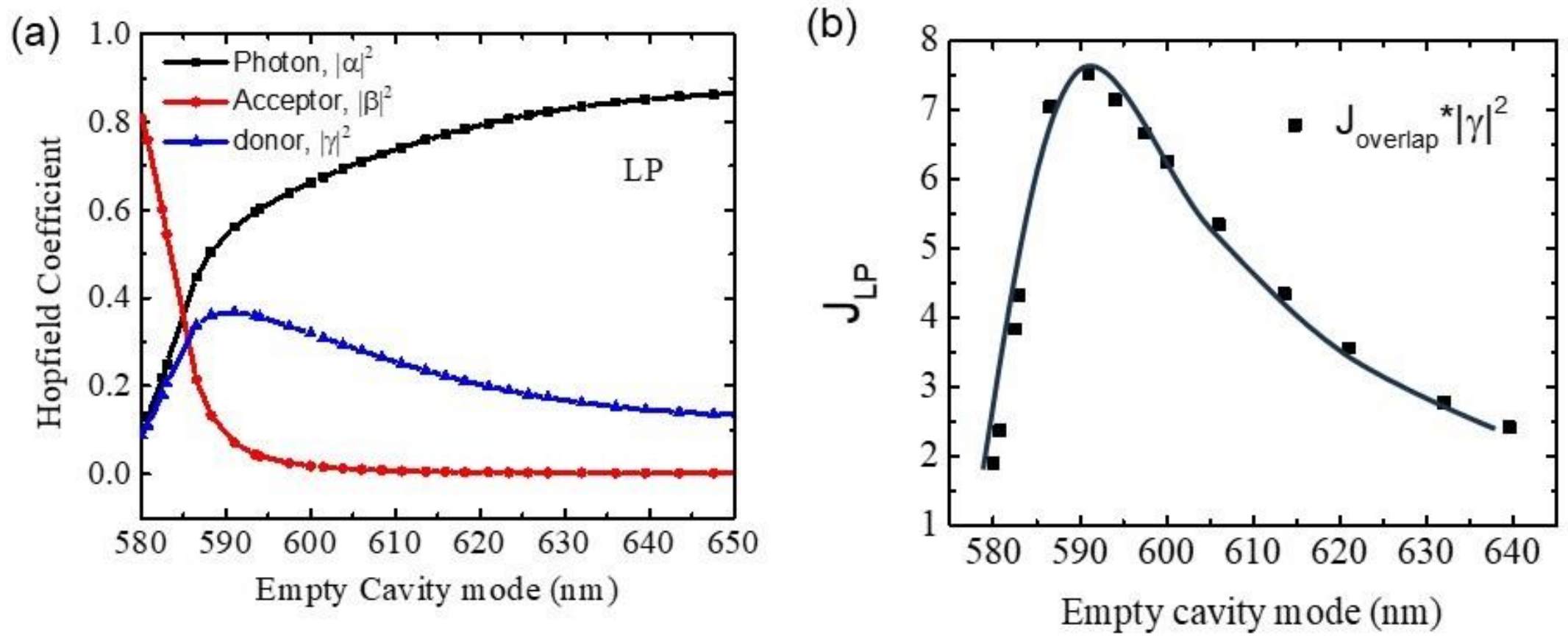

**Figure S11:** (a) Hopfield Coefficient for LP state (b) Absorption overlap of the LP state, uncoupled donor, and uncoupled acceptor plotted with respect to empty cavity mode position.

$\gamma_C$, $\gamma_A$, and $\gamma_D$ are the dephasing rates for cavity, acceptor and donor respectively, calculated from the cavity, acceptor and donor absorption FWHM respectively. where, $\gamma_C$ = 150 meV; $\gamma_A$ =40 meV; and $\gamma_D$ =35 meV. $g_A$ and $g_D$ are the coupling strength for acceptor and donor respectively. $g_D$=106 meV, $g_A$=55 meV. The eigenvalues of the above matrix gives the position of the polaritonic states, while the eigenvectors (α, β, γ) provide information about the composition of the cavity, acceptor, and donor components within the polaritonic states named as Hopfield coefficients.

## Section 14: Energy transfer probability:

We model the system using a time–dependent Schrödinger equation with a (time-independent) Hamiltonian that accounts for the interactions among: a cavity mode with energy $E_c$, the electronic excited state of a donor species with energy $E_D$ and that of an acceptor with energy $E_A$ (see section S13 above). Since donors and acceptors are placed at a separation distance larger than the Förster radius, we ignore possible direct donor–acceptor energy transfer.

We describe the stationary state of this three–component system with superposition states:

$$|\Psi\rangle = \alpha\left|D, A; 1_{\mathbf{k},p}\right\rangle + \beta|D^*, A; 0\rangle + \gamma|D, A^*; 0\rangle,$$

where the basis set describes states of the non–interacting system where the energy is in a cavity photon $\left|D, A; 1_{\mathbf{k},p}\right\rangle$ with wavevector $k$ and polarisation $p$, the electronic excited donor state $|D^*, A; 0\rangle$ and the electronic excited acceptor state $|D, A^*; 0\rangle$ respectively. The coefficients $\alpha, \beta, \gamma$ satisfy $|\alpha|^2 + |\beta|^2 + |\gamma|^2 = 1$, and are found as solutions to the equation

$$\hat{H}|\Psi\rangle = E|\Psi\rangle,$$

which also results in a set of three allowed energy values assigned to the lower ($E_L$), middle ($E_M$) and upper ($E_U$) polariton states, and three polariton states that we denote as $|LP\rangle$, $|MP\rangle$ and $|UP\rangle$.

As the system is interrogated by shining light onto it, we assume that at time $t = 0$ the state of the system is

$$|\Psi(0)\rangle = \left|D, A; 1_{\mathbf{k},p}\right\rangle,$$

which consists of both donor and acceptor species in their electronic ground states and one photon in a cavity mode. Temporal evolution of the system leads to a redistribution of energy and at some time $t$, the state of the system is described by

$$|\Psi(t)\rangle = e^{-i\hat{H}t/\hbar}|\Psi(0)\rangle.$$

Following Schäfer *et al*,[9] we quantify "energy transfer" by evaluating the projection of the state of the system at any point through its temporal evolution, on the "transfer target state", namely the one where the energy resides in the electronic excited state of $A$. More concisely, the probability of energy transfer is quantified by evaluating

$$P(t) = |\langle D, A^*; 0|\Psi(t)\rangle|^2,$$

a time-dependent function that exhibits Rabi oscillations as shown in **Figure S12a**

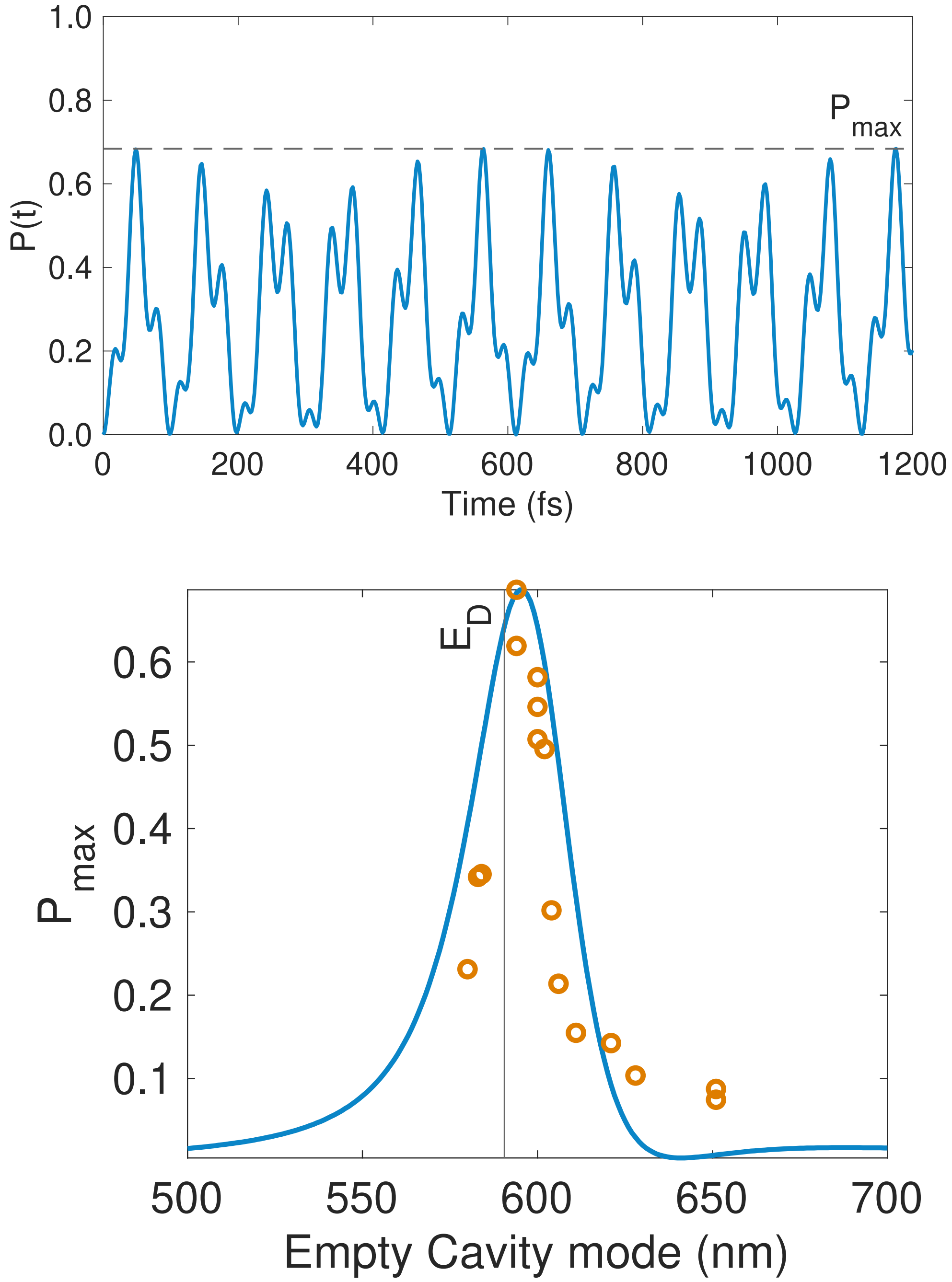

**Figure S12: (a)** Time evolution of P(t) exhibiting Rabi oscillations. The maximum value of P(t), $P_{max}$, is shown with a horizontal dashed line. **(b)** The maximum probability $P_{max}$ as a function of the energy (blue spectrum) of different cavity modes is plotted on top the experimental data points (orange circles) as given in Figure 4c

In the figure shown above we show the result of a numerical evaluation of the maxim of P(t) ($P_{max}$) for the case wherein $E_D$ = 2.100 eV, $E_A$ = 2.046 eV, $J_D$ = 105 meV and $J_A$ = 55 meV. In this figure we plot the maximum of $P(t)$ as a function of the energy of the cavity mode (shown in nanometers). The calculated spectrum closely resembles our results of **Figure 4c**, providing further evidence that strong light–matter coupling is responsible for the observed increase in photo-responsivity. We note that this model neglects effects of dissipation, which no doubt is present in our devices and include the low cavity Q achieved with a silicon mirror, in addition to possible radiative decay from the J-aggregates of TDBC. As discussed, transfer is further increased by achieving a stronger cavity-acceptor interaction.

****************